\documentclass[preprint,unsortedaddress,amsmath,amssymb,superscriptaddress]{revtex4-2}
\usepackage{epsf}
\usepackage{bm}
\usepackage{hhline}

\usepackage{amsfonts}
\usepackage{latexsym, mathtools}
\usepackage{graphicx}

\usepackage{newfloat}

\usepackage{hyperref}
\hypersetup{
    colorlinks=true, 
    linktoc=all,     
    linkcolor=blue,
    citecolor=blue,
    urlcolor=black,
    breaklinks=true
    }  
\usepackage[all]{hypcap}

\renewcommand{\baselinestretch}{1.3}

\begin{document}

\title{Direct observation of a spin-polarized surface Dirac gap in the antiferromagnetic topological insulator NdBi}

\author{Yuto~Fukushima}
\affiliation{Institute for Solid State Physics, The University of Tokyo, Kashiwa, Chiba 277-8581, Japan}

\author{Hiroyuki~S.~Suzuki}
\affiliation{Institute for Solid State Physics, The University of Tokyo, Kashiwa, Chiba 277-8581, Japan}

\author{Masayuki~Ochi}
\affiliation{Department of Physics, The University of Osaka, Toyonaka, Osaka 560-0043, Japan.}
\affiliation{Forefront Research Center, The University of Osaka, Toyonaka, Osaka 560-0043, Japan.}

\author{Takuma~Nishida}
\affiliation{Institute for Solid State Physics, The University of Tokyo, Kashiwa, Chiba 277-8581, Japan}

\author{Toshiya~Ikenobe}
\affiliation{Institute for Solid State Physics, The University of Tokyo, Kashiwa, Chiba 277-8581, Japan}

\author{Hiraku~Saito}
\affiliation{Institute for Solid State Physics, The University of Tokyo, Kashiwa, Chiba 277-8581, Japan}

\author{Yuto~Kinoshita}
\affiliation{Institute for Solid State Physics, The University of Tokyo, Kashiwa, Chiba 277-8581, Japan}

\author{Kaishu~Kawaguchi}
\affiliation{Institute for Solid State Physics, The University of Tokyo, Kashiwa, Chiba 277-8581, Japan}
\affiliation{Maryland Quantum Materials Center, Department of Physics, University of Maryland, College Park, MD 20742, USA}

\author{Hiroaki~Tanaka}
\affiliation{Institute for Solid State Physics, The University of Tokyo, Kashiwa, Chiba 277-8581, Japan}

\author{Ayumi~Harasawa}
\affiliation{Institute for Solid State Physics, The University of Tokyo, Kashiwa, Chiba 277-8581, Japan}

\author{Takushi~Iimori}
\affiliation{Institute for Solid State Physics, The University of Tokyo, Kashiwa, Chiba 277-8581, Japan}

\author{Koichiro~Yaji}
\affiliation{Center for Basic Research on Materials, National Institute for Materials Science (NIMS), 3-13 Sakura, Tsukuba, 305-0003, Ibaraki, Japan}
\affiliation{Unprecedented-Scale Data Analytics Center, Tohoku University, 468-1 Aoba, Aramaki-Aza, Aoba-Ku, Sendai, 980-8578, Japan}

\author{Masashi~Tokunaga}
\affiliation{Institute for Solid State Physics, The University of Tokyo, Kashiwa, Chiba 277-8581, Japan}

\author{Taro~Nakajima}
\affiliation{Institute for Solid State Physics, The University of Tokyo, Kashiwa, Chiba 277-8581, Japan}
\affiliation{Institute of Materials Structure Science, High Energy Accelerator Research Organization, Tsukuba, Ibaraki 305-0801, Japan}
\affiliation{RIKEN Center for Emergent Matter Science, Wako 351-0198, Japan}

\author{Kiyohisa~Tanaka}
\affiliation{SOKENDAI (The Graduate University for Advanced Studies), Okazaki 444-8585, Japan}
\affiliation{UVSOR Synchrotron Facility, Institute for Molecular Science, Okazaki 444-8585, Japan}

\author{Takahiro~Morimoto}
\affiliation{Department of Applied Physics, The University of Tokyo, Hongo, Tokyo, 113-8656, Japan}
\affiliation{Department of Physics, Kyoto University, Kyoto, 606-8502, Japan}

\author{Ryo~Mori}
\affiliation{Institute for Solid State Physics, The University of Tokyo, Kashiwa, Chiba 277-8581, Japan}
\affiliation{Maryland Quantum Materials Center, Department of Physics, University of Maryland, College Park, MD 20742, USA}

\author{Takeshi~Kondo}
\email{kondo1215@issp.u-tokyo.ac.jp}
\affiliation{Institute for Solid State Physics, The University of Tokyo, Kashiwa, Chiba 277-8581, Japan}
\affiliation{Trans-scale Quantum Science Institute, The University of Tokyo, Tokyo 113-0033, Japan}

\date{\today}

\maketitle

\textbf{A gapped Dirac cone in antiferromagnetic topological insulators (AFM-TIs) is essential for realizing the quantized anomalous Hall effect, yet its intrinsic magnitude remains controversial. In the prototypical AFM-TI MnBi$_2$Te$_4$, however, the observed Dirac gap is significantly smaller than theoretical predictions or nearly absent, highlighting a long-standing discrepancy. Here we demonstrate that NdBi hosts a spin-polarized topological surface Dirac cone with a sizable gap of $17 \pm 2$~meV at the Dirac point, in excellent agreement with density functional theory predictions (12~meV). 
This direct observation is enabled by high-resolution 7~eV laser-based spin- and angle-resolved photoemission spectroscopy, which unambiguously disentangles the spin-polarized surface Dirac cone from bulk-derived states. 
Furthermore, through temperature dependence and controlled surface contamination, we establish a direct linkage between surface magnetism and the Dirac gap, demonstrating that the gap opens only in the presence of surface magnetic order. 
These results provide definitive spectroscopic evidence for an intrinsic Dirac gap in AFM-TIs and establish NdBi as a model platform to resolve the long-standing discrepancy between experiment and theory in AFM-TIs.}
\\

\begin{center}
  \textbf{Introduction}
\end{center}

Topological insulators (TIs) realize a striking quantum state in which the bulk remains insulating while the surface hosts gapless Dirac fermions protected by time-reversal symmetry. Most TIs discovered to date have been realized in spin-orbit-coupled semiconductors with weak electronic correlations. More recently, strongly correlated TIs have attracted significant attention as a fertile ground for realizing exotic topological phases, including topological superconductors, topological Kondo insulators, and topological Mott insulators~\cite{Fu2008-ia, Dzero2010-jq,  Miyazaki2012-ll, Alexandrov2015-xg, Ohtsubo2019-pb, Weng2014-sm, Pesin2010-fh, YamajiUnknown-xz}.

Antiferromagnetic topological insulators (AFM-TIs) provide a compelling platform for exploring correlated topological phases. In AFM-TIs, it is predicted that long-range magnetic order opens an energy gap at the Dirac point of the topological surface state (Dirac gap), leading to the emergence of one-dimensional chiral edge states along the crystal hinges~\cite{Mong2010-os}. The MnBi$_{2n}$Te$_{3n+1}$ family, the first identified AFM-TI~\cite{Otrokov2019-lx}, has exhibited a variety of exotic responses, such as the quantum anomalous Hall effect and axion electrodynamics under magnetic fields~\cite{Chang2013-ik, Chang2015-qw,Deng2020-wd,Ge2020-xo,Liu2020-jp,Zhao2021-yc,Lin2022-an}. 

Despite these remarkable discoveries, a central challenge persists: the magnitude of the Dirac gap observed by angle-resolved photoemission spectroscopy (ARPES) is frequently inconsistent with theoretical predictions, and in many cases, absent~\cite{Zhang2019-kk,Li2019-gl,Otrokov2019-lx,Rienks2019-ho,Chen2019-qw,Hao2019-fo,Li2019-yv,Nevola2020-do,Swatek2020-my,Garnica2022-zn,Li2024-jy,Yang2024-cp}. Although early studies of MnBi$_2$Te$_{4}$ reported a sizable gap exceeding 50 meV~\cite{Otrokov2019-lx,Rienks2019-ho}, later investigations clarified that it primarily originates from bulk states, while the surface Dirac cone remains nearly gapless even under magnetic order~\cite{Chen2019-qw,Hao2019-fo,Li2019-yv,Nevola2020-do, Swatek2020-my}. Furthermore, atomic-scale disorder and surface reconstruction have been indicated to strongly suppress the surface Dirac gap, rendering it highly sensitive to sample quality~\cite{Garnica2022-zn,Tan2023-mv,Li2024-jy,Yang2024-cp}. As a result, conclusive ARPES evidence of an intrinsic surface Dirac gap in AFM-TIs has remained elusive, especially in MnBi$_2$Te$_4$. Such a discrepancy limits the high-temperature applicability of these materials and obscures the understanding of the intricate interplay 
between bulk magnetism and topological surface state (TSS), which is central to correlated topological physics.

Recently, the type-A antiferromagnet NdBi with the N\'eel temperature ($T_\mathrm{N}$) of 24~K has been proposed as a second AFM-TI, and its topological properties have been intensively investigated~\cite{Honma2023-sg, Honma2024-px}. NdBi belongs to a class of low-carrier-density semimetals with a band structure featuring multiple Dirac cones. The valence bands are mainly derived from Bi $6p$ orbitals, forming two distinct hole pockets. In contrast, the conduction bands originate from Nd $5d$ states and form electron pockets near the X points~\cite{Hasegawa1985-rb}. Due to strong spin-orbit coupling, the band gap narrows systematically from NdP to NdSb~\cite{Honma2023-sg,Kuroda2018-zw,Duan2018-gr,Lou2017-ll,Nayak2017-wg,Honma2024-px,Li2018-xg}, culminating in a band inversion in NdBi. This inversion underlies the topological phase transition in these materials.

The presence of an odd number of Dirac cones along the $\bar{\Gamma}$--$\bar{\rm M}$ direction---one at $\bar{\Gamma}$ and two at $\bar{\rm M}$~\cite{Honma2023-sg}---underscores the topological nontriviality of NdBi and motivates detailed spectroscopic investigations. Beyond topology, rare-earth monopnictides have drawn considerable attention due to their strongly correlated effects in the low-carrier-density regime. The small Fermi energy ($E_\mathrm{F}$) in these systems weakens Thomas-Fermi screening, enabling long-range Coulomb interactions. As a result, the coupling between itinerant and localized $f$ electrons becomes a defining factor in their physical properties~\cite{Kasuya1995-gw}; a notable example is CeSb, which exhibits complex magnetic transitions known as the ``devil’s staircase”~\cite{Rossat-Mignod1977-jy,Heer1979-ha,Rossat-Mignod1983-jg,Kuroda2020-ii}. The sister compound NdBi, therefore, provides an ideal platform for studying exotic AFM-TI phases driven cooperatively by magnetism and strong electron correlations in a low-carrier-density system.
However, the effect of electron correlation on the topological properties, including the TSS, remains unclear. 

A recent ARPES study on NdBi reported a sizable energy gap at the topological Dirac point as large as 125~meV below the N\'{e}el temperature~\cite{Honma2023-sg}. This feature was attributed to time-reversal symmetry breaking induced by out-of-plane ferromagnetic order in the top surface layer (corresponding to the antiferromagnetic phase, Domain-C). In that work~\cite{Honma2023-sg}, however, a large Dirac gap ($\sim$ 80~meV) is also reported even for the side surface (corresponding to the antiferromagnetic phase, Domain-A), where the Dirac point is expected to remain gapless due to symmetry protection. This ambiguity likely arises from the complex ARPES signals involving both surface and bulk states, reflecting a broader unresolved issue in AFM-TIs: the persistent discrepancy between experiment and theory, particularly regarding the nature and origin of Dirac gaps.
Spin-resolved experiments that isolate the topological surface state from bulk states are essential to resolve this issue and fully understand the antiferromagnetic topological properties of this system.

In this work, we overcome these experimental limitations by employing high-resolution laser-based ARPES combined with spin-resolved measurements, enabling clear separation of the TSS from bulk states and its direct comparison with \textit{ab initio} calculations. This approach provides evidence of the intrinsic surface Dirac gap of 17$\pm$2~meV in NdBi that originates from the topological surface state, in agreement with theoretical predictions (12~meV). Our results resolve the persistent discrepancy between experiment and theory in AFM-TIs and reveal pronounced band reconstruction in the antiferromagnetic phase, underscoring the essential role of inherent magnetism in stabilizing gapped surface states. Although much smaller than previously reported~\cite{Honma2023-sg}, the intrinsic Dirac gap we revealed in NdBi is still larger than the negligible gap in MnBi$_{2n}$Te$_{3n+1}$~\cite{Chen2019-qw,Hao2019-fo,Li2019-yv,Nevola2020-do, Swatek2020-my}. These results not only resolve a central puzzle in AFM-TI research but also establish NdBi as a promising platform for topological device engineering.
\\

\begin{center}
  \textbf{Results}
\end{center}

We begin by presenting, in Fig.~1, the electronic structures around $\bar{\Gamma}$ above (Fig.~1b-e) and below (Fig.~1f-i and Fig.~1j-m) $T_{\rm N}$ obtained by ARPES and density functional theory (DFT) calculations. Below $T_{\rm N}$, the magnetic moments of Nd~4$f$ electrons are ordered ferromagnetically within the (001) plane and stacked antiferromagnetically along the [001] direction [Fig.~1f and j]. Therefore, three magnetic domains are created with antiferromagnetic (AFM) moments aligned along the $a$, $b$, and $c$ axes, for which we label AFM-A, AFM-B, and AFM-C, respectively. These magnetic domains can be distinguished from the symmetry of the Fermi surface near the $\bar{\Gamma}$ point. 
To reveal the electronic structure of individual magnetic domains, we employed ARPES with a focused 7~eV laser beam ($\sim$50 $\mu$m in the spot size), which enables not only high-resolution measurements, but also magnetic-domain-selective observations by spatially scanning the small beam spot across the (001) cleavage surface.

The paramagnetic (PM) phase has three sheets of Fermi surfaces around $\bar{\Gamma}$ in the bulk state (Fig.~1c): one square extending in the $\bar{\Gamma}$--$\bar{\rm M}$ direction, another smaller square, and the smallest circular one at the center. Our DFT calculations (Fig.~1a,e) identify that these Fermi surfaces correspond to the heavy hole, light hole, and electron pocket, respectively. In Fig.~1d,e, ARPES intensity maps for energy dispersion and Fermi surface are compared with surface calculations, which contain both bulk states projected onto the crystal surface and intrinsic surface states. 
In ARPES spectra, bulk bands with strong $k_z$ dispersion appear as a broad continuum. This is due to $k_z$ broadening arising from the surface sensitivity of photoemission. The data and calculations consistently show highly dispersive Dirac-like dispersion characterized by sharp-peak spectra. Although this Dirac-like band has been interpreted as a topological surface Dirac cone in prior studies, its nature remains ambiguous and will be revisited in a later section.

In AFM-A (Fig.~1f), the magnetic moments are ordered in the [100] direction, and the Brillouin zone (BZ) is folded back along $k_x$, resulting in the appearance of two-fold Fermi surfaces in the bulk states [Fig.~1h,i]. In addition to the band folding effect, additional surface states with sharp spectra emerge and cross the Fermi level, consistent with previous reports. In the DFT calculation, two extra branches are seen around $-0.3$~eV [black arrow in Fig.~1i], which, however, were not observed by ARPES likely due to the matrix-element effect. It is worth noting that AFM-B shares the same electronic structure as AFM-A, but rotated by 90 degrees (see Supplementary Note 2).

In AFM-C [Fig.~1j], the magnetic moment is aligned perpendicular to the cleavage plane, causing the BZ to fold back along $k_z$. Unlike AFM-A and AFM-B, AFM-C retains $C_4$ symmetry over the cleavage plane even under the magnetic order. While the Fermi surface [Fig.~1l,m] hardly changes from the PM phase, two surface bands (SS1, SS2) emerge, as also reported in previous studies~\cite{Kushnirenko2023-gt}. Our DFT calculations reproduce these surface states, supporting that the observed domain is indeed the AFM-C domain.

The basic electronic structures presented above have been reported before~\cite{Schrunk2022-sv,Wang2023-xf,Kushnirenko2023-gt,Honma2023-sg,Yamamoto2025-ex}. The main results of this work are the observation of the Dirac gap and the intrinsic interplay between bulk and surface states.  
Before delving into the details presented in Figs. 3 and 4, we highlight a key challenge in Fig.~2: the difficulty in distinguishing surface and bulk states. Overcoming this difficulty is crucial for identifying a genuine TSS and, thus, the intrinsic Dirac gap in NdBi.

In ARPES research, surface states are typically assigned to states that show no signs of $k_z$ dispersion. 
We found that in NdBi, some of the bulk bands exhibit negligible $k_z$ dispersion, and significantly, the 2D-like bulk band displays an energy dispersion resembling a Dirac cone. To clarify this, in Fig.~2a, we calculated the bulk bands of PM phase, integrated them over the entire $k_z$. Although this calculation does not include surface states, the obtained results show Dirac-like bands consisting of sharp peak spectra.
For comparison, we performed the surface calculations in Fig.~2b, which not only exhibit the bulk band projected onto the surface but also surface states. Our results indicate that the topological surface Dirac cone emerges only in a narrow energy window within the band gaps of the Dirac-like bulk bands. 
Since the bulk band gap is only $\sim$ 13\,meV, standard ARPES with a typical energy resolution of $\sim$ 15 meV can hardly distinguish the genuine surface Dirac cone from the bulk bands. As simulated in Fig.~2f, the surface Dirac cone can be identified when the resolution reaches a level of 2.5 meV (as employed in this work). 

The above consideration suggests that unveiling the topological surface Dirac cone requires high-resolution measurements and the capability to distinguish between surface and bulk states, advancing beyond previous studies~\cite{Nayak2017-wg,Honma2023-sg,Honma2024-px}. Our laser-based ARPES technique provides sufficiently high-resolution ($\Delta E=2.5$~meV) to distinguish them (Fig.~2f). A further decisive approach is to use the spin-resolved ARPES. Because the TSS is spin-polarized, whereas the bulk states are not, spin-resolved ARPES could disentangle the complex interplay between surface and bulk states in NdBi.

In Fig.~3a, we show the high-resolution ARPES image obtained in the PM phase. A crossing point around $-0.27$~eV indicates the band as the Dirac cone. Nevertheless, the energy distribution curve (EDC) at $\bar{\Gamma}$ (red curve), overlaid in the image, exhibits a shoulder-like feature rather than a clean single peak. As discussed in Fig.~2, one needs to consider the contribution of Dirac-like bulk states to the data. 

To resolve this, we performed spin-ARPES and plotted in Fig.~3g the obtained spin-resolved EDCs at positive and negative momenta $+k$ and $-k$ (dotted lines in Fig.~3a) equidistant from $\bar{\Gamma}$. We found that the upper part of the Dirac cone exhibits spin flipping across $k_x=0$~\AA$^{-1}$. 
In contrast, the bottom part of the band does not exhibit clear spin flipping, as it is dominated by the bulk band of the light hole band, which contributes significantly to the spectral intensity. These observations conclude that a spin-polarized Dirac cone indeed exists. A more detailed discussion of the bulk-state contribution to the measured spin polarization is provided in Supplementary Note 9.

The spectral relationship between the surface Dirac cone and the bulk states in the AFM-C domain (Fig.~3d) is essentially the same as in the PM phase (Fig.~3a); while the upper bands correspond to the surface state, the lower bands consist of both surface and bulk states that can be distinguished only in a narrow energy-momentum window. However, a clear difference from the PM phase is observed in the shape of the surface band. Intriguingly, the upper Dirac cone, which was V-shaped in the PM phase, transformed into an M-shaped dispersion in the AFM-C domain. As revealed later (Fig.~4a), the M-shaped dispersion is the product of hybridization between the original V-shaped upper Dirac cone and a hole-like surface band, which emerges under antiferromagnetism. The features of these bands are further examined by spin-resolved ARPES in Fig.~3h. The upper cone clearly exhibits spin polarization that reverses across $\bar{\Gamma}$, while the lower cone, which almost overlaps with bulk states, is not clearly resolved. The magnitude of spin polarization in the upper cone is approximately 20\% and gradually decreases with increasing $k_x$, as reproduced by our spin-dependent DFT calculations [Fig.~3i]. This behavior can be attributed to the non-spin-polarized SS1, which hybridizes with the TSS, giving rise to the M-shaped dispersion.

Most importantly, we found a two-peak structure in the EDC at $\bar{\Gamma}$ (red curve overlaid in Fig.~3d), as marked by the black arrows, in addition to the intense peak of the bulk band. These two peaks indicate the opening of a gap at the Dirac point. The curvature plot in Fig.~3e offers a clearer view of the gap opening in the AFM-C. Compared with the PM phase (Fig.~3b), the lower Dirac cone becomes better discernible (Fig.~3e), as the bulk band shifts downward in the AFM-C domain. In these experiments, we used an ARPES analyzer with a deflector, which allows a precise determination of the $\bar{\Gamma}$ point. In addition, the low-energy photons of the 7~eV laser we use enable us to minimize the angle integration for each EDC detection, thus achieving measurements with high momentum resolution. Under this carefully controlled experimental setup, we consistently obtained the same gap size ($17\pm 2$~meV) from several different samples.

The gap size ($17\pm 2$~meV) identified here for the surface Dirac cone at $\bar{\Gamma}$ is an order of magnitude smaller than the previously reported value ($\sim$125~meV)~\cite{Honma2023-sg, Honma2024-px}. 
This discrepancy likely stems from misidentifying the Dirac-like bulk bands shown in Fig.~2 as surface states. This issue arises not only from limited energy resolution but also from matrix element effects, which are sensitive to the photon energy used in ARPES.
To examine this, we measured the photon-energy dependence of the band-dispersion signals in NdBi at a synchrotron facility specialized in low-photon-energy experiments (Fig.~4). At a low photon energy ($h\nu$) of 8~eV (Fig.~4c), close to $h\nu=7$~eV in our laser-ARPES, the surface states (TSS, SS1, and SS2) are clearly observed with high intensity.
In contrast, at a higher $h\nu$ of 10~eV (Fig.~4d), these surface states almost entirely vanish, and the spectral weight is dominated by bulk bands (Fig.~4a,b). 
This photon-energy dependence is a consequence of matrix-element effects, by which lower photon energies more effectively excite the TSS in NdBi. This implies that the large gap ($\sim$125~meV) reported in previous studies at $h\nu = $ 60~eV~\cite{Honma2023-sg, Honma2024-px} likely corresponds to a bulk band gap rather than the topological surface Dirac gap.

Notably, similar matrix-element effects have been reported in MnBi$_2$Te$_4$, which is well established as an AFM-TI~\cite{Hao2019-fo,Chen2019-qw,Baigeng_NC}. The TSS in that system is also clearly resolved only at low photon energies, whereas at higher photon energies (e.g., above $\sim$20~eV), the ARPES spectra are dominated by bulk bands. This can lead to an apparent large gap originating from bulk states being misinterpreted as the topological Dirac gap, even though the intrinsic TSS gap is much smaller.
The close analogy between MnBi$_2$Te$_4$ and NdBi underscores that low-$h\nu$ ARPES is very effective for directly accessing the intrinsic topological surface states in AFM-TIs.

Magnetic transitions in antiferromagnets are typically viewed as simple on-off switches between PM and ordered phases. The richness of their magnetic behavior can also be unveiled by the intermediate-temperature evolution. In particular, tracing how the M-shaped surface band (Fig.~3d) gradually develops as the interaction between the original TSS (the surface Dirac cone) and the emerging surface state below $T_{\rm N}$ offers new insight into the temperature-driven dynamic interplay between topology and magnetism. Our high-resolution ARPES measurements uniquely revealed this previously unexplored temperature evolution.

Figure~5a exhibits the temperature evolution of bands in a wider energy window for Fig.~3d. Two surface states (SS1 and SS2), which are absent in the PM phase, abruptly emerge just below $T_{\rm N}$. As the temperature decreases, both the surface states shift significantly upward in energy, while the energy position of the Dirac cone remains unchanged. In particular, SS1 with a hole-like dispersion intersects the upper Dirac cone, and the band hybridization occurs at these intersections, leading to the formation of the M-shaped dispersion. The magnitude of the hybridization gap ($\Delta_{\mathrm{Hyb}}$) increases as the temperature decreases. The gap opens isotropically, as shown in Fig.~5d, which presents the momentum-resolved gap distribution in the two-dimensional ($k_x, k_y$) plane at 21~K and 13~K. The complete temperature-dependent dataset from 25~K to 13~K in 1~K steps is provided as Supplementary Movie 1.

The band shift with temperature and the associated evolution of the hybridization gap are shown in Fig.~5b, which plots the band dispersions determined around the intersection between SS1 and the surface Dirac cone.
To establish a direct link between the hybridization gap and the magnitude of the magnetic order, we additionally performed neutron scattering measurements. In particular, we measured the temperature dependence of the integrated intensity of the 110 magnetic Bragg reflection $I_{110}$, whose square root $\sqrt{I_{110}}$ is proportional to the magnetic moment of Nd (Fig.~5c).
As demonstrated in Fig.~5d, we find excellent agreement between the temperature evolutions of $\sqrt{I_{110}}$ (reflecting the magnetic order parameter) and the experimentally observed hybridization gap $\Delta_{\mathrm{Hyb}}$. This provides compelling evidence that the gap opening originates from band hybridization induced by the exchange interaction associated with the magnetic order.
Neutron scattering is a bulk-sensitive probe. Therefore, the observation of a hybridization gap between the surface bands indicates that the surface magnetism evolves in tandem with the bulk magnetic order, at least on a freshly cleaved surface. This result also suggests that the hybridization gap provides a sensitive probe of the surface magnetism.

The temperature evolution of surface states below $T_{\rm N}$---specifically, the emergence and energy shift of SS1 and its hybridization with the topological Dirac cone---can be understood as follows. 
In NdBi, AFM ordering folds the bulk bands, and a bulk gap opens at the magnetic zone boundary where the original and folded bulk bands intersect and mutually hybridize via the exchange interaction. 
In AFM-C, the bulk band folding occurs along the $k_z$ direction, thereby opening the corresponding bulk gap around $\bar{\Gamma}$, which overlaps the original bulk gap in the PM phase. In general, surface states tend to appear inside bulk band gaps. This gap increases as temperature decreases, reflecting the growth of the magnetic moment and the concomitant enhancement of the exchange interaction. Likewise, SS1 and SS2 of NdBi emerge within this bulk gap that opens below $T_{\rm N}$. 
As the gap widens, SS1 shifts upward in energy. 
Concurrently, SS1 intersects the surface Dirac cone, thus a hybridization gap $\Delta_{\mathrm{Hyb}}$ of these surface bands opens to form the M-shape dispersion. This hybridization arises from exchange interactions driven by surface magnetism, which increases as the bulk magnetism intensifies at lower temperatures. 
The emergence of the M-shaped surface band in the AFM phase is reproduced by our DFT calculations (Supplementary Note 6). 

Our DFT calculations reveal a pronounced reconstruction of the bulk bands associated with the AFM phase transition (Supplementary Note 6), leading to a substantial enlargement of the bulk band gap. Importantly, this provides a natural explanation for the sizable gap of $\sim$125~meV reported for AFM-C domain in Ref.\cite{Honma2023-sg}. A similarly large gap is observed in our synchrotron-ARPES measurements, in which the spectra are dominated by bulk contributions due to matrix-element effects, thereby demonstrating that such large gaps originate from bulk bands rather than from topological surface states.

A large Dirac gap ($\sim$80~meV) was also reported at $\bar{\mathrm{M}}$ in AFM-A domain\cite{Honma2023-sg}, despite the theoretical expectation that the Dirac point remains gapless due to symmetry protection. According to our DFT calculations, this apparent inconsistency can be consistently resolved by recognizing that the reported gap does not reflect an intrinsic surface Dirac gap, but instead arises from bulk-band contributions (Supplementary Note 7).

Since the early days of topological insulator research, surface contamination has been widely employed to probe perturbative effects on TSS~\cite{Kong2011-yr,Benia2011-wj,Wang2012-ht,Yang2020-bc,Zhu2011-mb,King2011-cg,Valla2012-bo}. This approach is particularly important for AFM-TIs, as studies of the MnBi$_{2n}$Te$_{3n+1}$ family suggest that surface contamination may be a key factor underlying the conflicting reports on the presence or absence of a surface Dirac gap. In the case of NdBi, this method is expected to provide deeper insights, as the hybridization gap can serve as an indicator of surface magnetism. The relationship between the Dirac gap and surface magnetism can therefore be identified through simultaneous observation of the hybridization gap.

In our study, we exploited natural surface contamination. We found a drastic change in both the hybridization and Dirac gaps even under an ultra-high vacuum of $\sim 3\times 10^{-11}$ torr in the measurement chamber. 
Figure~5f displays the time evolution of the ARPES image for NdBi in the energy range capturing both the hybridization gap and the Dirac gap. Over time, the spectral quality, as evaluated by peak width, remained almost constant, indicating that surface degradation is not severe during the four-hour measurement.
We found that the energy position of SS1 ($E_{\rm{SS1}}$) remained unchanged over time.
However, the $\Delta_{\mathrm{Hyb}}$ vanished four hours after the sample was cleaved (black arrows in Fig.~5f). This is attributed to the loss of surface magnetism. Our results suggest that the surface magnetic order is more fragile than the surface electronic structure and can be suppressed even when the band dispersion remains well-defined, with the spectral sharpness, and hence the surface quality, nearly unchanged.
 
Very intriguingly, we found that the Dirac gap, clearly observed on a freshly cleaved surface, disappeared after three hours. This is evident in the EDC spectra at $\bar{\Gamma}$ (Fig.~5f): the double-peak structure associated with the gapped Dirac cone merges into a single peak after four hours, consistent with the Dirac gap closure. We emphasize again that the spectral peak width remained nearly constant throughout these measurements. This rules out surface-degradation-induced spectral broadening as the cause and indicates that these gap closures are intrinsic. The Dirac gap closure is coincident with the hybridization gap closure, thus validating that the Dirac gap originates from surface magnetism.

The time-dependent evolution of the band structure is schematically illustrated in Fig.~5g. The surface Dirac cone originates from the bulk band topology, as in the PM phase (left panel). The bulk magnetic order (or bulk exchange interactions) generates SS1, which intersects with the surface Dirac cone. On a freshly cleaved surface, the surface magnetic order aligns with the bulk magnetic order. The associated surface exchange interactions give rise to two types of gaps (middle panel): (i) a hybridization gap between the surface Dirac cone and SS1, resulting in an M-shaped surface band, and (ii) the Dirac gap, which opens due to time-reversal symmetry breaking induced by surface magnetism. Over time, residual gas gradually accumulates on the crystal surface. The surface Dirac cone and SS1 remain present since the surface contamination (right panel) does not affect either the bulk band topology or the bulk magnetism. Nonetheless, the magnetic order is lost at the surface, as supported by the disappearance of the hybridization gap $\Delta_{\mathrm{Hyb}}$, and the Dirac gap closes. 

Our DFT calculations reveal that surface states in the AFM-C domain are more localized on the crystal surface than those in the PM phase (Supplementary Note 8). While the decay length of the TSS into the bulk is approximately 17~nm in the PM state, it is reduced to 2-4~nm in the AFM-C domain. This reduction occurs because a large bulk gap, opened by bulk magnetism, significantly decouples the TSS from the bulk states in the AFM phase. When the TSS becomes more localized at the surface, it experiences the surface magnetism more strongly. Therefore, forming the gapped Dirac cone in NdBi requires both bulk magnetism, which decouples the TSS from the bulk, and surface magnetism, which couples directly to the localized TSS. Our experimental data in Fig.~5 indeed demonstrate that the Dirac gap is highly sensitive to the presence or absence of surface magnetic order, opening or closing accordingly.

\begin{center}
  \textbf{Discussion}
\end{center}

The observed exchange gap in the Dirac cone of NdBi ($17\pm 2$~meV) is in excellent agreement with our DFT calculations (12~meV). Importantly, examination of the hybridization gap $\Delta_{\mathrm{Hyb}}$ confirmed that this close match between theory and experiments occurs only when the surface magnetism is ideally aligned in the out-of-plane direction. Notably, the intrinsic gap we identified in NdBi is significantly larger than the vanishingly small gap observed in MnBi$_{2n}$Te$_{3n+1}$~\cite{Chen2019-qw,Hao2019-fo,Li2019-yv,Nevola2020-do,Swatek2020-my}. These results will provide an important clue toward the ultimate resolution of a central puzzle in AFM-TI research: the theory-reality mismatch.

In MnBi${2}$Te${4}$, a relatively small gap ($< 3$ meV) limits the observation of the quantum anomalous Hall effect (QAHE) to ultra-low temperatures ($\sim 1.4$~K)\cite{Deng2020-wd}. In contrast, the substantially larger gap observed in NdBi ($17\pm 2$~meV) suggests the possibility of realizing QAHE at elevated temperatures on the order of 10 K. Although further materials optimization will be required, this result points to a promising route toward higher-temperature QAHE.

We further note that thin-film fabrication offers a promising approach toward device realization. Quantum confinement in thin films is expected to drive a semimetal-to-semiconductor transition while preserving both magnetism and band topology~\cite{Ho2023-wt,Inoue2019-bl}. In addition, electrostatic gating could enable tunability of the chemical potential and enhance the effective gap, offering a viable strategy to optimize QAHE conditions.

Finally, we point out that previous studies on magnetic topological insulators in the MnBi$_{2n}$Te$_{3n+1}$ family treat magnetism as a perturbation that opens a Dirac gap to an effective Hamiltonian that describes nonmagnetic states~\cite{Zhang2019-kk,Nevola2020-do,Yan2021-vv}. In contrast, our study reveals that magnetism in NdBi intricately intertwines with both bulk and surface states. Such magnetism is beyond the scope of a simple perturbative framework for understanding the magnetic topological states. The need to go beyond the perturbative regime arises from the strong interplay between localized 4$f$ electrons and itinerant carriers---a qualitatively different situation from that in MnBi$_{2n}$Te$_{3n+1}$, where magnetic properties are perturbatively induced by magnetic layers composed of Mn atoms with 3$d$ orbitals that are located in proximity to topological Bi$_2$Te$_3$ layers. These findings for the second-generation AFM-TI (NdBi) mark a significant step beyond the first-generation AFM-TI (MnBi$_{2n}$Te$_{3n+1}$) and provide valuable guidance for the future design and discovery of antiferromagnetic topological insulators among a wide variety of magnets.

\begin{center}
  \textbf{Methods}
\end{center}

\noindent\textbf{Calculations}\\
First-principles calculations were performed using the WIEN2k software~\cite{Blaha2018-zf,Blaha2020-tl}.
We used the modified Becke--Johnson (BJ) potential~\cite{Tran2009-ps}.
The maximum modulus for the reciprocal lattice vectors $K_{\mathrm{max}}$ was chosen so that $RK_{\mathrm{max}}=8$ where the muffin-tin radius $R$ for Nd and Bi atoms was 2.5 Bohr.
We used a $19\times19\times19$ $\bm{k}$-mesh with the inclusion of the spin-orbit coupling.
The experimental lattice constant $a=6.41$ \AA \ was taken from Ref.~\onlinecite{Abdusalyamova1988-zf}.

For the paramagnetic phase, we adopt the open-core treatment for the Nd-$f$ orbitals.
For the spin-polarized phases, we adopt the DFT$+U$ method~\cite{Anisimov1993-hn, Liechtenstein1995-ki} with $U=6.3$ eV and $J=0.7$ eV for the Nd-$f$ orbitals as adopted in the previous study~\cite{Wang2023-xf}.
We constructed the maximally localized Wannier functions~\cite{Marzari1997-dz, Souza2001-vg} for the Nd-$d$ and Bi-$p$ orbitals from the calculated band structure using the Wannier90 software~\cite{Pizzi2020-sm}.
The semi-infinite-slab tight-binding models constructed from these Wannier functions were used for calculating the surface spectra in the way described in Ref.~\onlinecite{Lopez_Sancho1985-gi}.\\

\noindent\textbf{Single crystal growth}\\
NdBi single crystals were grown by the Bridgman method with a sealed tungsten crucible and a high-frequency induction furnace. High-purity Nd (5N) and Bi (6N) metals with the respective composition ratios were used as starting materials. The obtained samples were characterized by X-ray diffraction patterns.\\

\noindent\textbf{ARPES measurements}\\
All the high-resolution laser-ARPES measurements were performed at the Institute for Solid State Physics, the University of Tokyo, using two different setups with 6.994~eV lasers~\cite{Shimojima2015-ox,Yaji2016-ts}. The domain-selective measurements (Fig.~1) and the observation of the Dirac cone (Fig.~3) were performed using an ARPES system equipped with a ScientaOmicron DA30L spectrometer. Spin-resolved ARPES (SARPES) measurements were conducted with VLEED-type spin-detectors installed on the ScientaOmicron DA30L spectrometer. We used an effective Sherman function of $S_{eff}=0.27$. The VLEED-type spin detector operates by periodically switching the target magnetization during data acquisition, which effectively removes instrumental asymmetry.  The energy (angular) resolution was set to 2.5~meV (0.1$^\circ$) and 21~meV (1.4$^\circ$) for ARPES and SARPES, respectively. 

To investigate the temperature dependence of spectra in detail, we employed another ARPES system equipped with a Scienta R4000 analyzer. The excellent temperature tunability in the sample manipulator enabled temperature scans in fine steps of 1~K while maintaining the deviation within $\pm$0.1~K~\cite{Kuroda2020-ii}. The scans were first performed from lower to higher temperatures, followed by another temperature scan from higher to lower temperatures to ensure data reliability. 

For temperature-dependent measurements, the samples were cleaved at the lowest temperature (13 K), whereas for other measurements, cleavage was performed at the measurement temperature. The beam spot size was approximately 50~$\mu$m, which is significantly smaller than the typical magnetic domain size ($>200$~$\mu$m, Supplementary Note 2), ensuring that the measurements predominantly probe a single magnetic domain. In addition, a total of 38 samples were cleaved and measured for ARPES. Although the data quality varies from sample to sample, the main results presented in the manuscript were found to be highly reproducible. We note that NdBi has a rock-salt crystal structure, for which a single crystallographic termination is expected upon cleavage.

The synchrotron-ARPES experiments with tunable low photon energies were performed at the BL7U beamline of UVSOR, equipped with an MBS A-1 analyzer. The energy resolution was better than 15~meV, and the angular resolution was 0.3$^\circ$. The synchrotron data (Fig.~4 and Fig.~S5) exhibit a slight distortion in the band shape due to a minor imperfection in the analyzer lens setting. While this affects the precise determination of the band dispersion and prevents a fair comparison across different photon energies, it does not affect the key observation that the surface-state intensity is strongly suppressed at 10 eV due to matrix element effects.

In all these ARPES experiments, we used the linear polarization with the electric field aligned in the incident plane. The crystals were cleaved $in~situ$ at the measurement temperatures. The base pressure during the ARPES measurements was better than 1$\times$10$^{-8}$~Pa.\\

\noindent\textbf{Neutron scattering experiment}\\
Neutron scattering experiments were conducted using the POlarized Neutron Triple-Axis triple-axis spectrometer PONTA installed at the 5G beamhole in the Japan Research Reactor 3 (JRR-3) of Japan Atomic Energy Agency~\cite{Nakajima2024-iz}. The spectrometer was operated in a two-axis diffraction mode. An incident neutron beam with the energy of 34.05~meV was obtained using a pyrolytic graphite monochromator. A single crystal sample with dimensions of 22.52~mm$^3$ was oriented in the (H,K,0) horizontal scattering plane and loaded in a clamp-type uniaxial-stress cell~\cite{Aso2005-dm,Nakajima2011-of}. A uniaxial stress of 25~MPa was applied onto the (001) surfaces of the sample, so that the magnetic domains with the magnetic moments parallel to the [001] direction were selected at low temperatures. The cell was attached to the cold head of a closed-cycle $^4$He refrigerator and cooled to the lowest temperature of 2.3~K. Further experimental details are provided in Supplementary Note 4.



\begin{center}
  \textbf{Acknowledgements}
\end{center}

The sample preparation in this work was carried out using a glovebox in the facilities of the Materials Design and Characterization Laboratory at the Institute for Solid State Physics, the University of Tokyo. A part of this work was conducted at the BL7U of UVSOR Synchrotron Facility, Institute for Molecular Science (IMS program 24IMS6656 and 25IMS6663). The neutron scattering experiment at PONTA in JRR-3 was carried out along the proposals (No. 24401), and was partly supported by the Neutron Science Laboratory of the Institute for Solid State Physics, the University of Tokyo.\\

\begin{center}
  \textbf{Fundings}
\end{center}

R.M. discloses support for the research of this work from The University of Tokyo Edge Capital Partners (UTEC)-University of Tokyo Future Society Initiative (UTokyo FSI) Research Grant Program, and JST PRESTO Grant No. JPMJPR24LA. T.K. discloses support for the research of this work from the JSPS KAKENHI [Grant Numbers: JP25H01250, JP25H01246, and JP26H02014], MEXT Q-LEAP [Grant Number: JPMXS0118068681], The Asahi Glass Foundation, The Murata Science Foundation, The Mitsubishi Foundation, and Toray Science Foundation.\\




\newpage
\renewcommand{\baselinestretch}{1.1}
%
\begin{figure*}[t]
  \includegraphics[width=6.6in]{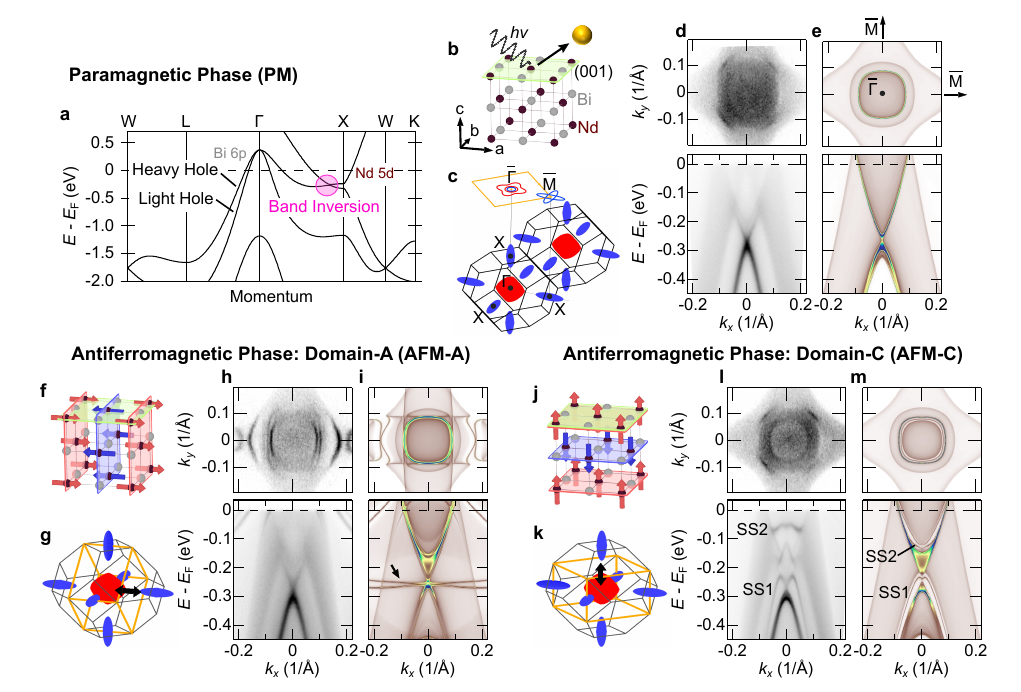}
  \caption{\textbf{Electronic structure of NdBi across magnetic phases and domains.}
\textbf{a}, Bulk band structure of NdBi in the paramagnetic (PM) phase, highlighting the band inversion between Bi~6$p$ and Nd~5$d$ orbitals.
\textbf{b}, Crystal structure of NdBi, with the cleavage plane shown in green.
\textbf{c}, Schematic of the bulk and surface Brillouin zones (BZs).
\textbf{d}, Fermi surface map and ARPES intensity plot of the band dispersion along $\bar{\Gamma}$--$\bar{\rm M}$ of NdBi measured by ARPES in the paramagnetic phase. 
\textbf{e},  Fermi surface and band dispersion along $\bar{\Gamma}$--$\bar{\rm M}$ calculated using density functional theory (DFT).
\textbf{f}, Magnetic structure of the AFM-A domain.
\textbf{g}, Folded BZ due to magnetic ordering along [100].
\textbf{h},  Fermi surface map and band dispersion along $\bar{\Gamma}$--$\bar{\rm M}$ measured by ARPES.
\textbf{i}, Corresponding DFT results.
\textbf{j-m}, Same as (f-i), but for the AFM-C domain.
Yellow lines in the (g) and (k) represent the folded BZ.
Two surface states (SS1 and SS2) are resolved in the AFM-C phase (l,m).}
\end{figure*}
%

%
\begin{figure}[t]
  \includegraphics[width=6in]{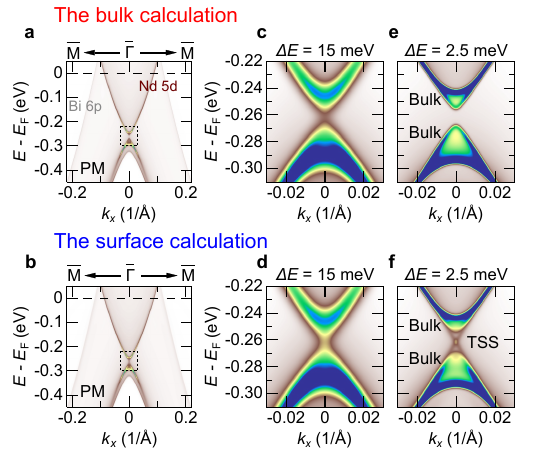}
  \caption{
\textbf{Energy resolution effect on the bulk band and the surface Dirac cone in NdBi.}
\textbf{a},\textbf{b}, DFT-calculated band dispersion of paramagnetic phase along the $\bar{\Gamma}$--$\bar{\rm M}$ direction, showing the spectral weight of bulk (\textbf{a}) and surface (\textbf{b}) states.
\textbf{c-d}, Simulated ARPES images showing the impact of energy resolution on the visibility of the bulk band gap marked by a dotted rectangle.
The calculations in \textbf{a} and \textbf{b} are convoluted with energy resolutions of 15~meV in \textbf{c} and \textbf{d}, and 2.5~meV in \textbf{e} and \textbf{f}, respectively.
The topological surface state (TSS) is only clearly resolved in the surface calculation when using high energy resolution (2.5~meV), as shown in \textbf{f}.}
\end{figure}

%
\begin{figure}[t]
  \includegraphics[width=4.3in]{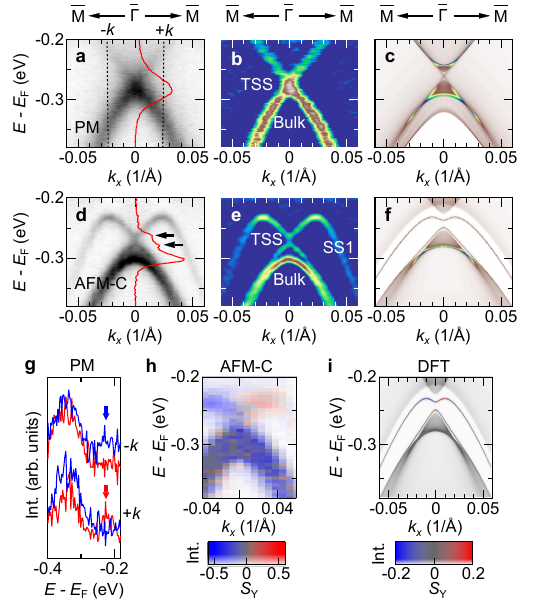}
  \caption{
\textbf{Observation of gapped surface Dirac cone entangled from the bulk band.}
\textbf{a}, High-resolution ARPES image along the $k_x$ direction ($k_y = 0$~\AA$^{-1}$) measured in the paramagnetic phase ($T = 44$~K). The energy distribution curve (EDC) at $\bar{\Gamma}$ is overlaid in red.
\textbf{b}, Curvature plot of the ARPES data shown in \textbf{a}.
\textbf{c}, DFT-calculated band structure in the paramagnetic phase.
\textbf{d-f}, Same as \textbf{a-c} but for the AFM-C domain ($T = 13$~K), showing the emergence of an M-shaped surface band and the opening of a Dirac gap.
The EDC in \textbf{d} shows two peaks at $\bar{\Gamma}$ (black arrows), indicating the gap.
\textbf{g}, Spin-resolved EDC spectra at $+k$ and $-k$ positions (as indicated in \textbf{a}), showing opposite spin polarizations. Red and blue lines represent up and down spin components, respectively.
\textbf{h}, Spin-resolved band dispersion measured in the AFM-C domain.
\textbf{i}, Corresponding spin-polarized DFT calculation, confirming the spin texture of the gapped surface Dirac cone.
  }
\end{figure}
%
%

\begin{figure}[t]
\includegraphics[width=4in]{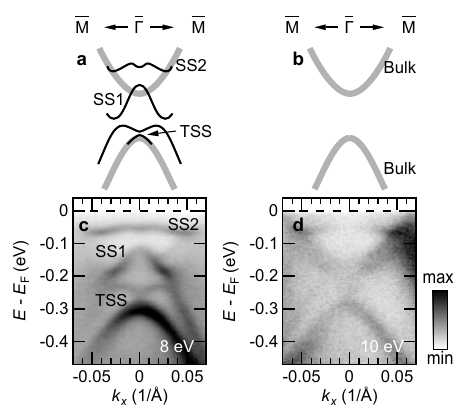}
\caption{
\textbf{Suppression of surface state ARPES signals at higher photon energies.}
\textbf{a,b} Schematic band dispersions representing both surface states (TSS, SS1, and SS2) and bulk states (\textbf{a}), and only bulk states (\textbf{b}), corresponding to \textbf{c} and \textbf{d}, respectively.
\textbf{c,d} ARPES intensity maps along the $\bar{\Gamma}$--$\bar{\mathrm{M}}$ direction measured at photon energies of 8~eV (\textbf{c}) and 10~eV (\textbf{d}).
While both surface and bulk bands are observed at the lower photon energy (8~eV), matrix-element effects strongly suppress the surface-state intensity at the higher photon energy (10~eV), dominantly leaving only the signals from the bulk states, as illustrated in \textbf{a} and \textbf{b}, respectively.
}
\end{figure}

%
\begin{figure*}[t]
  \includegraphics[width=6.1in]{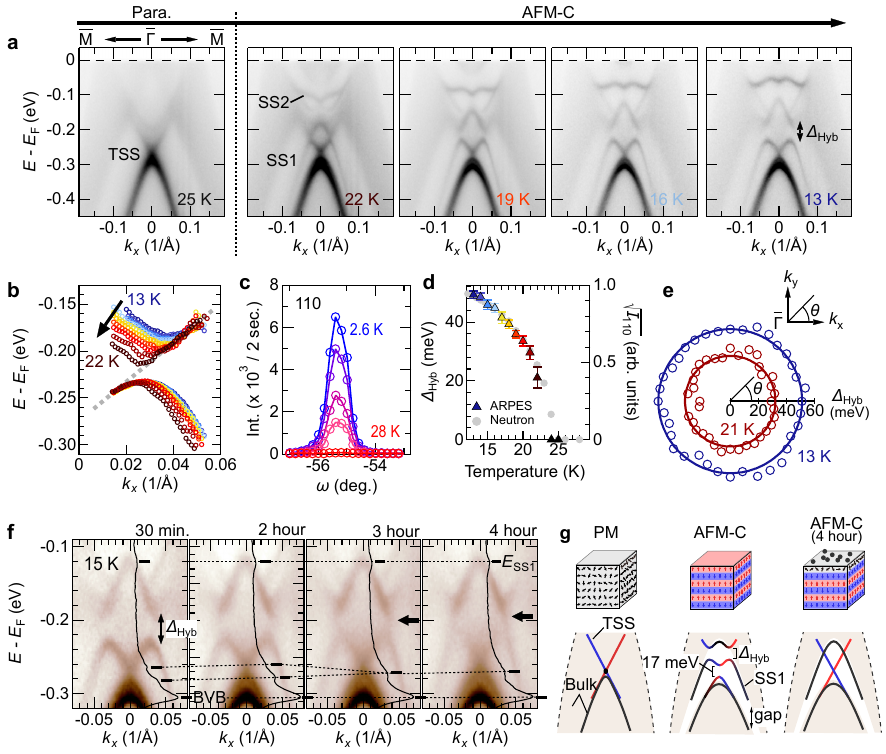}
  \caption{
\textbf{Evolution of the hybridization gap and Dirac cone with temperature and surface contamination.}
\textbf{a}, Temperature evolution of ARPES images from 25~K (paramagnetic phase) to 13~K (AFM-C domain), showing the emergence of surface states (SS1 and SS2) and the development of the hybridization gap ($\Delta_\mathrm{Hyb}$).
\textbf{b}, Temperature dependence of the Dirac cone and SS1, determined from the EDC peak positions. For the EDC fittings, we employed a Gaussian function combined with a linear background. No constraints were imposed on the fitting parameters. The gray dotted line traces the TSS in the PM phase.
\textbf{c}, Temperature dependence of the 110 magnetic reflection peak measured by neutron scattering. The horizontal axis $\omega$ denotes the sample rotation angle.
\textbf{d}, Comparison of the temperature dependence of the hybridization gap observed by ARPES ($\Delta_\mathrm{Hyb}$, left) and the square root of the integrated intensity of the 110 magnetic Bragg reflection ($\sqrt{I_{110}}$, right). $\Delta_{\mathrm{Hyb}}$ is defined as the energy difference between the top and bottom of the parabolic bands in \textbf{b}.
\textbf{e}, Angular dependence of $\Delta_\mathrm{Hyb}$ measured at 21~K and 13~K. The angle $\theta$ is defined from the $\bar{\Gamma}$--$\bar{\rm M}$ direction in the ($k_x$, $k_y$) plane, as illustrated in the inset.
\textbf{f}, ARPES images measured at 15 K at different time intervals after cleavage, showing the degradation of the Dirac and hybridization gaps due to surface contamination. EDCs at $\bar{\Gamma}$ are overlaid.
\textbf{g}, Schematic illustration of the electronic structure in the paramagnetic phase, in the AFM-C domain immediately after cleavage, and in the AFM-C domain after 4 hours, highlighting the evolution of the surface Dirac gap (17~meV), hybridization gap ($\Delta_\mathrm{Hyb}$), and surface state SS1.
  }
\end{figure*}

\end{document}


\title{Supplementary Information for \\``Direct observation of a spin-polarized surface Dirac gap in the antiferromagnetic topological insulator NdBi"}

\author{Yuto~Fukushima}
\affiliation{Institute for Solid State Physics, The University of Tokyo, Kashiwa, Chiba 277-8581, Japan}

\author{Hiroyuki~S.~Suzuki}
\affiliation{Institute for Solid State Physics, The University of Tokyo, Kashiwa, Chiba 277-8581, Japan}

\author{Masayuki~Ochi}
\affiliation{Department of Physics, Osaka University, Toyonaka, Osaka 560-0043, Japan.}
\affiliation{Forefront Research Center, Osaka University, Toyonaka, Osaka 560-0043, Japan.}

\author{Takuma~Nishida}
\affiliation{Institute for Solid State Physics, The University of Tokyo, Kashiwa, Chiba 277-8581, Japan}

\author{Toshiya~Ikenobe}
\affiliation{Institute for Solid State Physics, The University of Tokyo, Kashiwa, Chiba 277-8581, Japan}

\author{Hiraku~Saito}
\affiliation{Institute for Solid State Physics, The University of Tokyo, Kashiwa, Chiba 277-8581, Japan}

\author{Yuto~Kinoshita}
\affiliation{Institute for Solid State Physics, The University of Tokyo, Kashiwa, Chiba 277-8581, Japan}

\author{Kaishu~Kawaguchi}
\affiliation{Institute for Solid State Physics, The University of Tokyo, Kashiwa, Chiba 277-8581, Japan}

\author{Hiroaki~Tanaka}
\affiliation{Institute for Solid State Physics, The University of Tokyo, Kashiwa, Chiba 277-8581, Japan}

\author{Ayumi~Harasawa}
\affiliation{Institute for Solid State Physics, The University of Tokyo, Kashiwa, Chiba 277-8581, Japan}

\author{Takushi~Iimori}
\affiliation{Institute for Solid State Physics, The University of Tokyo, Kashiwa, Chiba 277-8581, Japan}

\author{Koichiro~Yaji}
\affiliation{Center for Basic Research on Materials, National Institute for Materials Science (NIMS), 3-13 Sakura, Tsukuba, 305-0003, Ibaraki, Japan}
\affiliation{Unprecedented-Scale Data Analytics Center, Tohoku University, 468-1 Aoba, Aramaki-Aza, Aoba-Ku, Sendai, 980-8578, Japan}

\author{Masashi~Tokunaga}
\affiliation{Institute for Solid State Physics, The University of Tokyo, Kashiwa, Chiba 277-8581, Japan}

\author{Taro~Nakajima}
\affiliation{Institute for Solid State Physics, The University of Tokyo, Kashiwa, Chiba 277-8581, Japan}
\affiliation{Institute of Materials Structure Science, High Energy Accelerator Research Organization, Tsukuba, Ibaraki 305-0801, Japan}
\affiliation{RIKEN Center for Emergent Matter Science, Wako 351-0198, Japan}

\author{Kiyohisa~Tanaka}
\affiliation{SOKENDAI (The Graduate University for Advanced Studies), Okazaki 444-8585, Japan}
\affiliation{UVSOR Synchrotron Facility, Institute for Molecular Science, Okazaki 444-8585, Japan}

\author{Takahiro~Morimoto}
\affiliation{Department of Applied Physics, The University of Tokyo, Hongo, Tokyo, 113-8656, Japan}

\author{Ryo~Mori}
\affiliation{Institute for Solid State Physics, The University of Tokyo, Kashiwa, Chiba 277-8581, Japan}
\affiliation{Maryland Quantum Materials Center, Department of Physics, University of Maryland, College Park, MD 20742}

\author{Takeshi~Kondo}
\email{kondo1215@issp.u-tokyo.ac.jp}
\affiliation{Institute for Solid State Physics, The University of Tokyo, Kashiwa, Chiba 277-8581, Japan}
\affiliation{Trans-scale Quantum Science Institute, The University of Tokyo, Tokyo 113-0033, Japan}

\date{\today}

\maketitle

{\noindent\textbf {Supplementary Note 1: Sample characteristics}}
\vspace{0.3\baselineskip}

\begin{figure*}[h]
  \centering
  \includegraphics[width=0.9\textwidth]{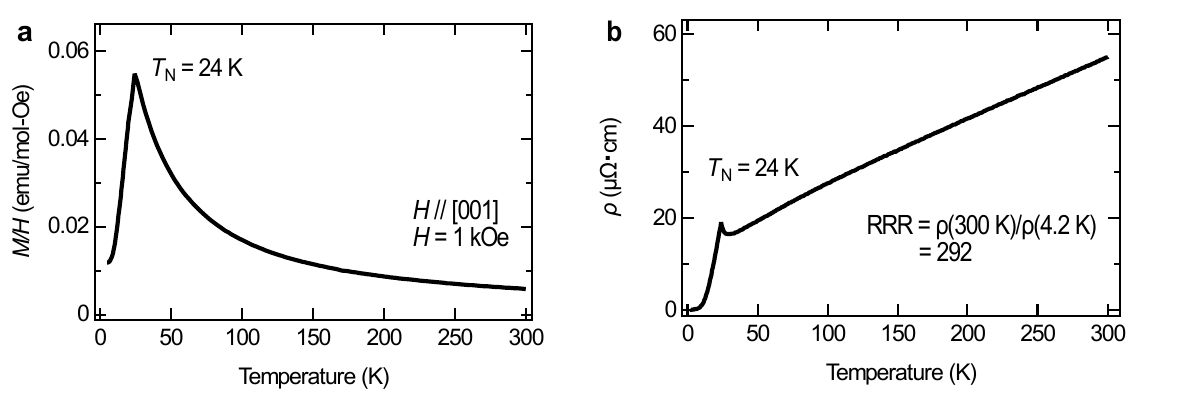}
  \label{sup_magnetization}
  \caption{
\textbf{a}, Temperature dependence of the magnetization $M/H$ of NdBi, measured with an applied field of 1~kOe along the [001] direction.
\textbf{b},  Temperature dependence of the electrical resistivity $\rho$ measured in zero field. The N\'eel temperature $T_\mathrm{N} = 24$~K is clearly observed in both panels. The residual resistivity ratio is estimated to be $\mathrm{RRR} = \rho(300~\mathrm{K})/\rho(4.2~\mathrm{K}) = 292$, indicating high crystal quality.}
\end{figure*}

The sample quality was confirmed by measurements of magnetic susceptibility (Fig.~\ref{sup_magnetization}a) and electrical resistivity (Fig.~\ref{sup_magnetization}b). Magnetization was measured in a field of 1~kOe. Both measurements exhibit an anomaly at $T_{\mathrm{N}} = 24$~K. The residual resistivity ratio (RRR), estimated as $\rho(300\,\mathrm{K})/\rho(4.2\,\mathrm{K}) = 292$, indicates excellent crystal quality.\\

{\noindent\textbf {Supplementary Note 2: Optical microscopy image}}
\vspace{0.3\baselineskip}

The type-A antiferromagnetic order in a cubic lattice makes three types of magnetic domains with the ordered moments along [100], [010], and [001]. Following the notation in the main text, we refer to these magnetic domains (A, B, and C), respectively. To visualize the multi-domain state in our high-quality sample, we employed polarized optical microscopy, a technique previously applied to this~\cite{Honma2023-sg,Yamamoto2025-ex} and its sister compound, CeSb~\cite{Kuroda2020-ii,Arai2021-hn}. The measurement was performed at ISSP, the University of Tokyo. The details of the experimental setup are described in Ref.~\cite{Katakura2010-ym}. The sample was cleaved in the atmosphere and installed immediately into the vacuum chamber. 

\begin{figure*}[h]
  \centering
  \includegraphics[width=0.9\textwidth]{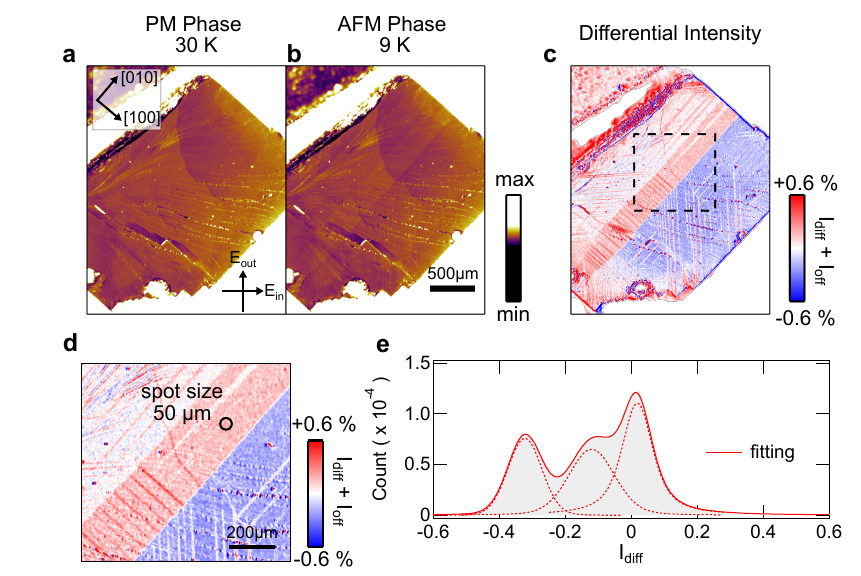}
    \label{sup_microscope}
  \caption{
Polarized optical microscope images taken at $T = 30$~K (paramagnetic phase: PM, \textbf{a}) and at $T = 9$~K (antiferromagnetic phase: AFM, \textbf{b}). We used the crossed Nicol configuration with the incident light polarized along the [110] direction to visualize the ordered magnetic domain. The differential intensity map $I_{\rm diff}$ obtained by subtracting \textbf{b} from \textbf{a} is presented in \textbf{c} using a color scale with a constant offset $I_{\rm off}=0.12$. \textbf{d},The magnified differential intensity map indicated by a rectangle in \textbf{c}. The spot size of our laser system, used for domain-selective ARPES measurements, is represented by an open circle. \textbf{e}, Distribution of differential intensities of \textbf{d}. The offset $I_{\rm off}$ for the color scale is determined from the middle peak of the fitting result.
}
\end{figure*}

The microscope image observed in the Antiferromagnetic (AFM) phase (Fig.~\ref{sup_microscope}a, $9$~K) shows additional contrast compared to the image observed in the paramagnetic (PM) phase (Fig.~\ref{sup_microscope}b, $30$~K). The temperature dependence is well visualized in the differential intensity map $I_{\rm diff}=(I_{\rm 9K} - I_{\rm 30K})\,/\,2(I_{\rm 9K} + I_{\rm 30K})\times 100$ (Fig.~\ref{sup_microscope}c). The change in intensities between before and after the magnetic transition is primarily due to the optical birefringence under the tetragonal magnetic symmetry in the AFM phase. The magnified differential intensity map indicated by the rectangle in Fig.~\ref{sup_microscope}c is shown in Fig.~\ref{sup_microscope}d. It clearly shows three domain regions represented by red, blue, and white areas. Domain sizes are basically larger than our laser spot ($\sim50~\mu$m) installed in our high-resolution laser-ARPES, enabling the domain-selective ARPES measurements.

For better visualization, the color scale has an offset of 0.12\%, determined from a distribution of the differential intensities in Fig.~\ref{sup_microscope}d (Fig.~\ref{sup_microscope}e). It shows three peaks attributed to the three magnetic domains. We chose the middle peak as the offset for the color scales used in Fig.~\ref{sup_microscope}c and d.\\

{\noindent\textbf {Supplementary Note 3: ARPES results for all three magnetic Domains}}
\vspace{0.3\baselineskip}

\begin{figure*}[h]
  \centering
  \includegraphics[width=0.9\textwidth]{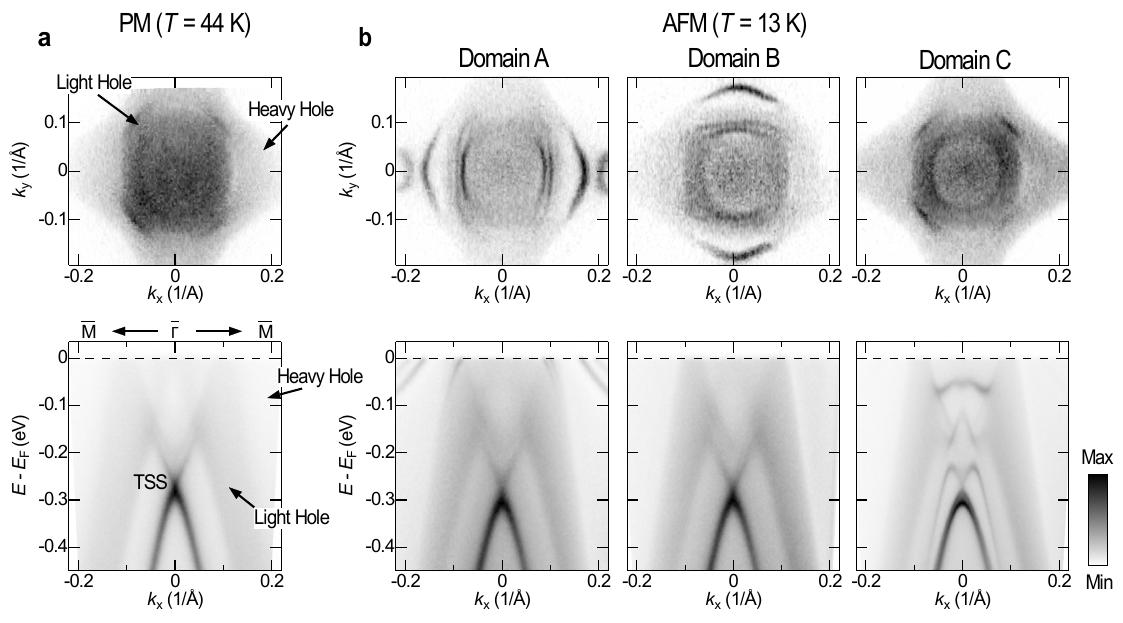}
    \label{sup_Domains}
  \caption{
  Fermi surface mappings (top panels) and ARPES band dispersion images along the $\overline{\Gamma}$-$\overline{\mathrm{M}}$ direction (bottom panels) measured at $T = 44$~K (paramagnetic phase: PM, \textbf{a}) and at $T = 13$~K (antiferromagnetic phase: AFM, \textbf{b}) for the three magnetic domains (A-C). In the PM phase, distinct light-hole and heavy-hole bands are resolved, and the topological surface state (TSS) appears within the bulk band gap. Upon cooling into the AFM phase, the Fermi surface and band structure exhibit strong domain dependence, reflecting the different magnetic arrangements.}
\end{figure*}

While the main text presents ARPES results only for Domains A and C, here we also include data for Domain B for comparison. Figure~S3a shows the Fermi surface and band dispersion maps measured in the paramagnetic phase (at $T$ = 44 K, Fig.~\ref{sup_Domains}a). Upon cooling the sample to $T$ = 13 K (below $T_\mathrm{N}$), these features drastically change due to the magnetic phase transition (Fig.~\ref{sup_Domains}b). 

Although a direct real-space domain map is not available in the present ARPES measurements, and the precise domain population has not been determined, we performed spatially resolved ARPES measurements by scanning the beam across the cleaved sample surface. This approach allows us to probe different regions of the sample and distinguish magnetic domains based on their Fermi surface characteristics.

We obtained ARPES data from all three domains on the same cleavage plane, as displayed in Fig.~S3b. The magnetic domains were identified based on distinct symmetry fingerprints of the Fermi surface: Domains A and B exhibit twofold symmetry, whereas Domain C displays fourfold symmetry. The domain assignment was performed by directly matching the measured Fermi surface patterns to these symmetry characteristics, providing an objective and unambiguous classification of the domains.\\

{\noindent\textbf {Supplementary Note 4: Characterization of magnetic domains under uniaxial stress via neutron scattering}}

\vspace{0.3\baselineskip}

\begin{figure*}[h]
  \centering
  \includegraphics[width=0.9\textwidth]{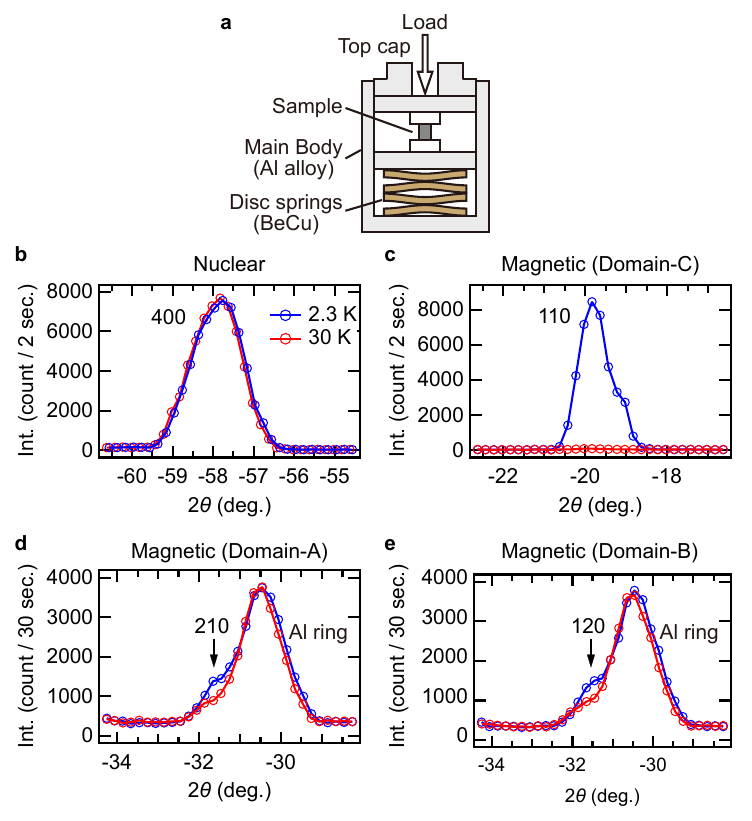}
    \label{sup_NeutronScattering}
  \caption{
\textbf{Temperature dependence of nuclear and magnetic Bragg reflections under uniaxial stress}  
\textbf{a}, Schematic of the uniaxial stress cell. The main body is constructed from an Al alloy, and the internal disc springs are made of BeCu. Uniaxial stress of 25~MPa is maintained by clamping the top cap.
\textbf{b}, 400 nuclear Bragg reflection.  
\textbf{c}, 110 magnetic reflection from the AFM-C domain.
\textbf{d,e}, 210 and 120 reflections corresponding to AFM-A and AFM-B, respectively. In panels \textbf{b}--\textbf{e}, data are shown for temperatures above $T_{\rm N}$ (30~K, red) and below $T_{\rm N}$ (2.3~K, blue). Black arrows denote magnetic peaks, while the temperature-independent background signals originate from the Al-alloy cell.
}
\end{figure*}

For the neutron scattering measurements, the single crystal sample was mounted in the uniaxial-stress cell with the (H,K,0) horizontal scattering plane (Figs.~\ref{sup_NeutronScattering}a). The (001) surface of the sample was polished to minimize possible inhomogeneity of the applied stress. The cell was clamped under a pressure of 25~MPa at room temperature before being installed into the vacuum chamber of the cryostat.

Fundamental neutron scattering results are presented in Figs.~\ref{sup_NeutronScattering}b--e. The 400 nuclear Bragg peak shown in Fig.~\ref{sup_NeutronScattering}b serves as a reference; its intensity remains constant across the antiferromagnetic (AFM) transition. As shown in Fig.~\ref{sup_NeutronScattering}c, we observed a strong magnetic Bragg peak at the reciprocal lattice point of (1,1,0), where the nuclear Bragg reflections is forbidden, at low temperatures. This reflection corresponds to the AFM-C domain, and disappeared in the paramagnetic phase at 30~K, demonstrating its magnetic origin.

To evaluate domain distribution under strain, we measured the 210 and 120 reflections corresponding to the AFM-A and AFM-B domains, respectively (Figs. \ref{sup_NeutronScattering}d, e). Besides the
magnetic peaks (black arrows), these spectra include temperature-independent signals from the Al cell. We measured integrated intensities of the 110, 210 and 120 reflections by $\theta - 2\theta$\ scans at low temperatures, and also repeated the same scans at high temperatures to subtract background signals. Taking into account the magnetic form factor of Nd$^{3+}$ and Lorentz factor, we estimated the volume fractions of the three magnetic domains from the integrated intensities, revealing that the AFM-C population exceeds 98~\%. This demonstrates that uniaxial strain selectively stabilizes a single magnetic domain.\\

{\noindent\textbf {Supplementary Note 5: Suppression of surface state spectra at higher photon energies.}}
\vspace{0.3\baselineskip}

In the main text, we used a 7~eV laser as the photon source for our ARPES measurements and revealed the surface state that emerges within the bulk band gap. Here, we demonstrate that the surface-state signals weaken at higher photon energies, such as 10~eV, and the ARPES spectra eventually become dominated by bulk band signals. For this purpose, we used synchrotron radiation tunable from 7~eV to higher photon energies at the BL07 of the UVSOR Synchrotron Facility.

\begin{figure*}[h]
  \centering
  \includegraphics[width=0.9\textwidth]{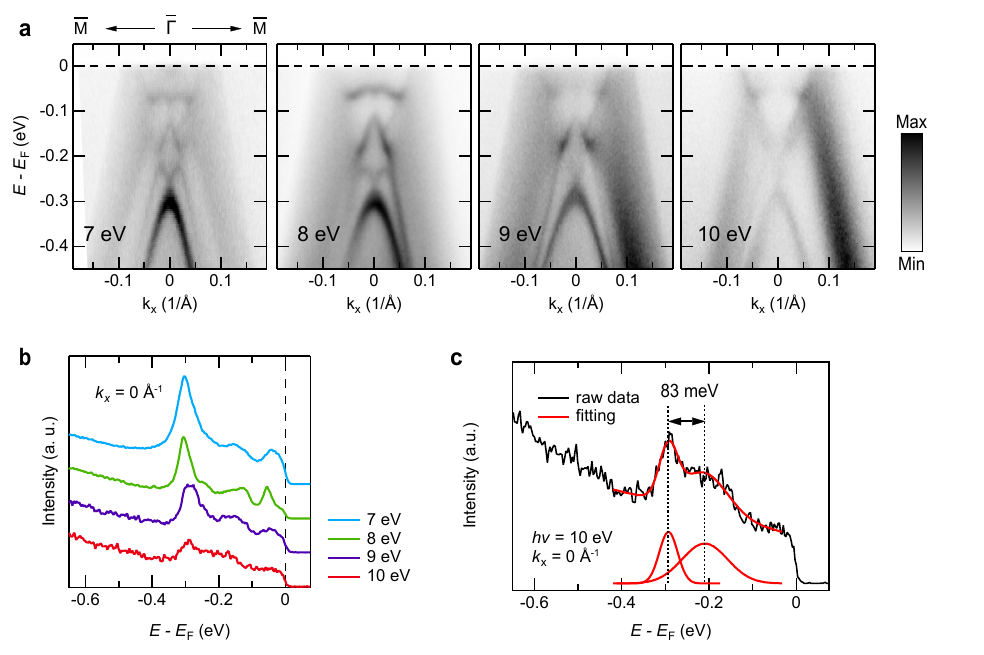}
    \label{sup_hvDepend}
  \caption{
\textbf{Photon-energy-dependent ARPES spectra measured along the $\overline{\Gamma}$-$\overline{\mathrm{M}}$ direction in the antiferromagnetic Domain C (AFM-C).}
\textbf{a}, ARPES band dispersions recorded with photon energies of 7, 8, 9, and 10~eV. The TSS are clearly visible at 7~eV but become increasingly suppressed with higher photon energy.  
\textbf{b}, Corresponding energy distribution curves (EDCs) at $k_x = 0$~\AA$^{-1}$ extracted from the data in (a), highlighting the spectral evolution with photon energy.  
\textbf{c}, EDC at $k_x = 0$~\AA$^{-1}$ measured at $h\nu = 10$~eV (black), fitted with two Gaussian peaks (red), yielding an energy separation of 83~meV. This large gap originates from the bulk band and not the Dirac surface state, as discussed in the text.
}
\end{figure*}

Figure~\ref{sup_hvDepend}a shows ARPES band dispersions along the $\bar{\Gamma}$-$\bar{\rm M}$ direction, obtained from the antiferromagnetic domain C (AFM-C). The left-to-right panel corresponds to data measured at photon energies from 7~eV to 10~eV. Below 9~eV, the two surface states (SS1 and SS2) are clearly observed. However, they almost vanish at 10~eV, most likely due to matrix element effects in photoemission. This is more clearly demonstrated in Fig.~\ref{sup_hvDepend}b by extracting the EDCs at $k_x = 0$~\AA$^{-1}$. The spectral shape varies with increasing photon energy, and at 10~eV, the in-gap surface state is not distinguishable from the bulk spectra. In Fig.~\ref{sup_hvDepend}c, the EDC at 10~eV was fitted with two Gaussian curves, yielding a gap magnitude of 83~meV. This is much larger than the Dirac gap (17~meV) determined from the data measured with a 7~eV laser ARPES in the main text.
The large gap (83~meV) does not originate from the Dirac surface state, but rather from the bulk band. This discrepancy, coming from the photon energies used, may be a primary reason that the energy gap claimed as the one for the Dirac point varies across different research groups. Note that a similar situation has been reported for the MnBi$_2$Te$_4$ system~\cite{Chen2019-qw,Baigeng_NC,Hao2019-fo}, in which photon energies lower than $\sim$18 eV are required to detect the surface Dirac state which emerges within the bulk band gap.\\

{\noindent\textbf {Supplementary Note 6: Substantial enhancement of the bulk band gap under antiferromagnetism}}
\vspace{0.3\baselineskip}

Here, we use band calculations around $\bar{\mathrm{\Gamma}}$ to support the argument made in Supplementary Note 5 that the large gap (83 meV) observed in the antiferromagnetic phase of NdBi originates from the bulk bands. In particular, we demonstrate that the calculated bulk bands with a Dirac-like shape significantly enhance the bulk gap in the antiferromagnetic phase, and the spectral gap observed by ARPES with a finite energy resolution could be as large as $\sim100$~meV. This is nearly one order of magnitude larger than the gap (12 meV) that opens at the Dirac point of the topological surface state (TSS) in the surface calculations.

\begin{figure*}[h]
  \centering
  \includegraphics[width=0.9\textwidth]{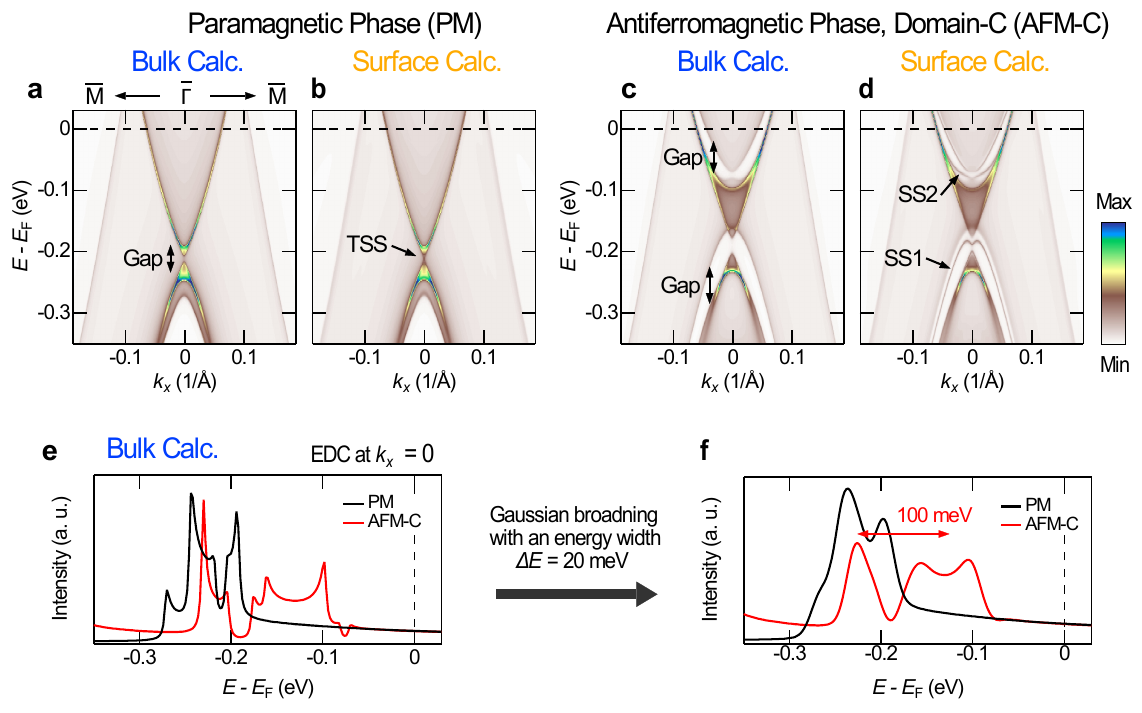}
    \label{sup_AFcalc}
  \caption{
\textbf{DFT band structure calculations along the $\overline{\Gamma}$-$\overline{\mathrm{M}}$ direction for NdBi in the paramagnetic phase (PM) and antiferromagnetic phase, Domain-C (AFM-C).}
\textbf{a}, The bulk band calculations that include only the bulk spectral weight in the PM phase. A small bulk gap is observed near the Dirac point.  
\textbf{b}, The surface band calculations that include both bulk and surface spectral weight in the PM phase, showing the emergence of a TSS within the bulk gap.  
\textbf{c}, The bulk band calculations, same as \textbf{a}, but for the AFM-C phase. Two distinct bulk gaps open due to antiferromagnetism.  
\textbf{d}, The surface band calculations, same as \textbf{b}, but for the AFM-C phase. Two surface states (SS1 and SS2) appear within the bulk gaps.  
\textbf{e}, Energy distribution curves (EDCs) at $k_x = 0$~\AA extracted from the bulk spectra in \textbf{a} and \textbf{c}. Black and red lines correspond to the PM and AFM-C phases, respectively.  
\textbf{f}, Simulated EDCs obtained by convolving the spectra in \textbf{e} with a Gaussian of full width at half maximum (FWHM) of 20~meV. The AFM-C phase exhibits an enhanced gap of $\sim$100~meV, consistent with the bulk gap observed in high-photon-energy ARPES in Fig.~\ref{sup_hvDepend}. }
\end{figure*}

As a first step, we reproduce in Fig.~\ref{sup_AFcalc}a,b the band calculations for the paramagnetic phase presented in the main Fig.~2. Figure~\ref{sup_AFcalc}a shows the ``bulk calculations" that include only the bulk bands, whereas Figure~\ref{sup_AFcalc}b shows the ``surface calculations" that include both the bulk and surface bands. The bulk bands (integrated over the entire $k_{\rm z}$) resemble a Dirac dispersion, characterized by sharp spectral features. A band gap opens between the bulk bands of the heavy hole pocket and the electron pocket, as indicated by an arrow in Fig.~\ref{sup_AFcalc}a. Within this bulk band gap, the TSS emerges, as revealed by the surface-inclusive calculation in Fig.~\ref{sup_AFcalc}b.

We next performed calculations for the antiferromagnetic (AFM) phase. Here, we selected the Domain-C (AFM-C) configuration and obtained results for both the bulk bands only (Fig.~\ref{sup_AFcalc}c) and the bulk plus surface states (Fig.~\ref{sup_AFcalc}d). This comparison clearly distinguishes bulk from surface bands. In AFM-C, the AFM order causes the heavy-hole and electron pockets to hold along the $k_{\rm z}$ direction, leading to additional band hybridizations and the opening of two band gaps (arrows in Fig.~\ref{sup_AFcalc}c). Within these newly formed gaps, two surface states (SS1 and SS2) appear, as shown in Fig.~\ref{sup_AFcalc}d. The lower-energy gap overlaps with the original bulk gap in the PM phase (Fig.~\ref{sup_AFcalc}a), within which a surface Dirac cone resides. This gap is enhanced by the exchange interaction of antiferromagnetism. Inside the enlarged bulk gap, the surface band (SS1) forms an M-shaped dispersion (Fig.~\ref{sup_AFcalc}d). As discussed in the main text, time-reversal symmetry breaking on the crystal surface opens an energy gap at the Dirac point of this M-shaped band. Although the gap is relatively small (12~meV), it agrees well with our ARPES results (17~meV).

As discussed in Supplementary Note 5, the intensity of surface state signals can become weak at high photon energies due to matrix element effects. Such data could be highly misleading, since the bulk band retains its Dirac-like shape, and the bulk band gap becomes enlarged in the antiferromagnetic phase. To simulate this situation, we extracted energy distribution curves (EDCs) at $\bar{\Gamma}$ from the bulk calculations without surface states (Fig.~\ref{sup_AFcalc}e) for both the paramagnetic phase (black curve) and the antiferromagnetic phase (red curve). To make the simulation more realistic, we introduced Gaussian broadening with an assumed energy resolution of 20~meV to the EDC in Fig.~\ref{sup_AFcalc}(f). The resulting EDCs indicate that the bulk band gap enlarged under antiferromagnetism reaches as $\sim$100~meV (red arrow). 

These arguments indicate that the estimation of the intrinsic topological Dirac gap requires carefully optimized experimental conditions: proper low photon energies and high resolution. Our experimental setting matches that condition. Moreover, our spin-resolved ARPES measurements are essential for distinguishing the bulk and surface states, which become the key to achieving our research goals. We emphasize that our experimental data are the first that decisively identifies the intrinsic topological Dirac gap of NdBi experimentally.\\

{\noindent\textbf {Supplementary Note 7: Electronic band evolution at the $\bar{\mathrm{M}}$ point across the antiferromagnetic phase transition investigated by DFT calculations}}
\vspace{0.3\baselineskip}

\begin{figure*}[h]
  \centering
  \includegraphics[width=0.9\textwidth]{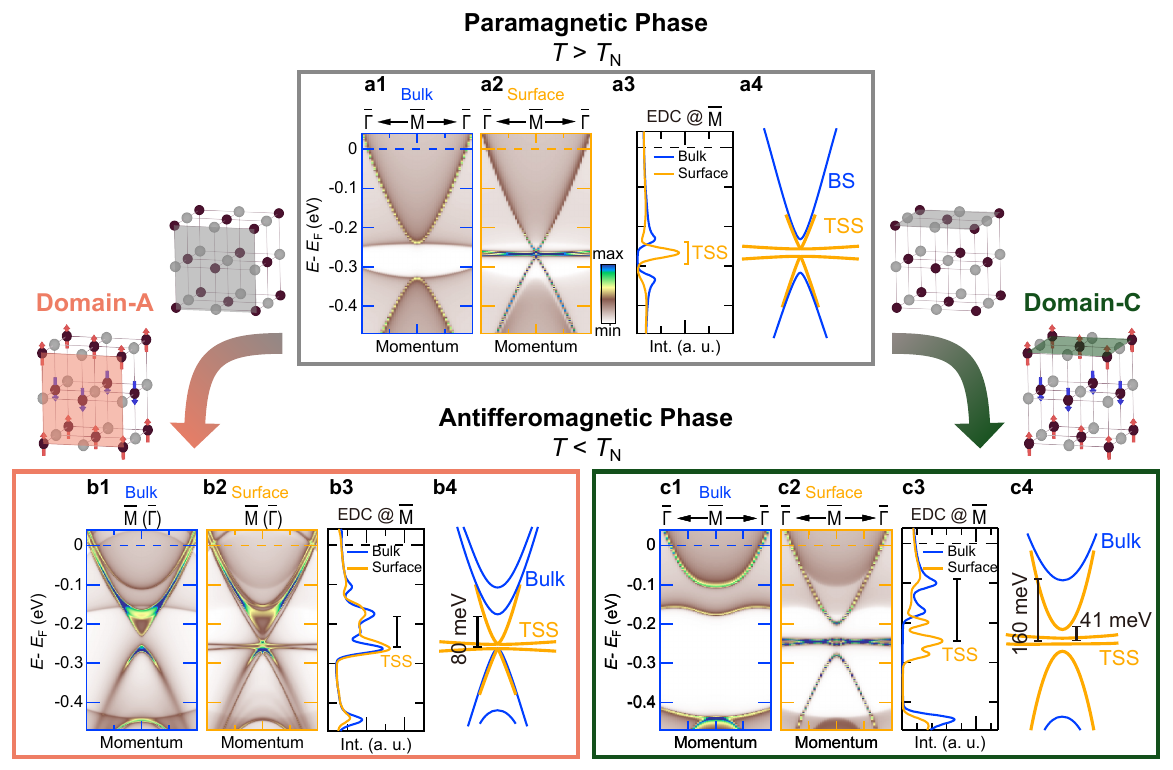}
    \label{sup_calcMbar}
  \caption{
\textbf{Calculated electronic structures around the $\bar{\mathrm{M}}$ point on the (001) surface.}
\textbf{a1-a4}, Calculations for the paramagnetic phase. 
\textbf{b1-b4}, Calculations for the antiferromagnetic phase, Domain-A (AFM-A). 
\textbf{c1-c4}, Calculations for the antiferromagnetic phase, Domain-C (AFM-C). 
\textbf{a1,b1,c1}, Calculated bulk spectral weights.
\textbf{a2,b2,c2}, Calculated surface spectral weights.
\textbf{a3,b3,c3}, Energy distribution curves (EDCs) at the $\bar{\mathrm{M}}$ point extracted from the surface (orange) and bulk (blue) calculations. Energy resolution of 20 meV is applied to simulate a realistic spectrum. 
\textbf{a4,b4,c4}, Schematics of the band structures. Blue lines represent bulk states and orange lines represent topological surface states (TSS). In c4, the large energy separation involving bulk states ($\sim$160 meV) and the intrinsic magnetic gap of the TSS (41 meV) are indicated.
   }
\end{figure*}

In the main text, we focus on the band structure around the $\bar{\mathrm{\Gamma}}$ point. 
Here, we investigate the band structures around the $\bar{\mathrm{M}}$, in particular, their evolution across the antiferromagnetic phase transition, using our DFT calculations (Fig.~\ref{sup_calcMbar}). 
The reliability of these calculations is supported by their excellent agreement with our ARPES results around the $\bar{\mathrm{\Gamma}}$ point. 

Before discussing this momentum region in detail, we note that the electronic bands around the $\bar{\mathrm{M}}$ do not directly reflect the underlying nontrivial $Z_2$ topology.

In NdBi, a single Dirac cone is located at the $\bar{\mathrm{\Gamma}}$ point, where it represents an odd number of topological surface states (TSSs) required by the nontrivial $Z_2$ invariant. While TSSs are also observed at the $\bar{\mathrm{M}}$ point, these arise from the projection of multiple bulk band inversions and give rise to an even number of Dirac cones. Such states appearing at the $\bar{\mathrm{M}}$ point may not be topologically protected; therefore, Dirac cones at the $\bar{\mathrm{M}}$ are not suitable for investigating the interaction between magnetism and the nontrivial $Z_2$ topology. This possibility is also mentioned in~\cite{Honma2023-sg}.

Measurements at the $\bar{\mathrm{\Gamma}}$ point, where the relevant TSS resides, are experimentally challenging due to $k_z$ broadening, which leads to overlap with projected bulk bands. Nevertheless, direct observation of a Dirac gap for the topological Dirac cone at the $\bar{\mathrm{\Gamma}}$ point remains indispensable for identifying the antiferromagnetic topological phase in this system.

Because of this intrinsic difficulty, the identification cannot rely solely on energy-momentum intensity maps and requires going beyond conventional ARPES approaches. This difficulty arises because bulk-derived Dirac-like bands and $k_z$ broadening fundamentally obscure the surface contribution in conventional ARPES measurements.

In this work, we combine high-resolution 7 eV laser ARPES with spin-resolved measurements. Spin-resolved ARPES is essential to distinguish the spin-polarized topological surface state from bulk-derived bands. This allows a direct identification of the TSS. Importantly, the gap opening at the Dirac point of the spin-polarized surface band is in quantitative agreement with our DFT surface calculations, further supporting the intrinsic origin of the observed Dirac gap. Taken together, the consistency between spin-resolved ARPES and surface-state calculations provides a compelling and self-consistent identification of the topological Dirac gap.

Still, momentum space around the $\bar{\mathrm{M}}$ point, where surface and bulk states are more clearly separated, can be advantageous for evaluating the impact of magnetism on the opening of a Dirac gap in NdBi associated with time-reversal symmetry breaking. The 7 eV laser ARPES employed in this work does not access this momentum region due to its low photon energy. 
Nevertheless, our DFT calculations allow us to investigate the evolution of the band structure across the antiferromagnetic transition. Importantly, they identify the relationship between bulk and surface bands and  
provide a consistent interpretation of previously reported large gaps~\cite{Honma2023-sg}, attributing them to bulk-band features rather than to the intrinsic TSS.
In particular, we find that sharp dispersive features can arise from both surface and projected bulk bands, and their distinction cannot be made based solely on spectral sharpness. Furthermore, we note that a large gap has been reported even in symmetry-protected situations where the Dirac point is expected to remain gapless~\cite{Honma2023-sg}. Within our framework, this apparent discrepancy can be naturally explained by bulk-band contributions rather than by the intrinsic gap opening of the TSS. Projected bulk bands can exhibit sharp dispersions and overlap with surface states, leading to an overestimation of the surface-state gap if not properly distinguished.

The above-mentioned mechanism is summarized in Fig.~\ref{sup_calcMbar}, where we present DFT calculations around the $\bar{\mathrm{M}}$ point above and below $T_\mathrm{N}$ (top and bottom panels, respectively) for both Domain-A and Domain-B (left and right panels, respectively). 

\vspace{0.5\baselineskip}
\noindent \textbf{Domain C:}

Figures~\ref{sup_calcMbar}a1,a2 show the spectral intensity maps of bulk and surface calculations in the paramagnetic phase above $T_\mathrm{N}$, while the corresponding results in the antiferromagnetic phase below $T_\mathrm{N}$ are shown in Figs.~\ref{sup_calcMbar}c1,c2. Upon entering the AF phase, the bulk gap becomes significantly enlarged.

In contrast, the intrinsic gap at the Dirac point of the TSS at the $\bar{\mathrm{M}}$ point remains much smaller, about $41$~meV (Fig.~\ref{sup_calcMbar}b5). Notably, the energy separation between the nearly flat TSS branch and the bulk bands is approximately $160$~meV (Fig.~\ref{sup_calcMbar}c4), which is comparable to the gap size reported as the TSS gap in Ref.~\cite{Honma2023-sg}.

This comparison indicates that the previously reported large gap can be attributed to the energetic separation between surface and bulk states, rather than to the intrinsic magnetic gap of the TSS. This result highlights the importance of reliably distinguishing surface and bulk contributions, as achieved here using spin-resolved ARPES.

\vspace{0.5\baselineskip}
\noindent \textbf{Domain A:}

Ref.~\cite{Honma2023-sg} also reported a similarly large gap at the Dirac point of the TSS for Domain A, where the Dirac point is expected to remain symmetry-protected and gapless. In Ref.~\cite{Honma2023-sg}, the origin of this large gap was not identified. Within our framework, however, this observation can be consistently explained by bulk-band contributions rather than by an intrinsic gap opening of the TSS Dirac point. Supporting DFT calculations are presented in Fig.~\ref{sup_calcMbar}b.\vspace{0.5\baselineskip}

The large gap reported in Ref.~\cite{Honma2023-sg} is likely influenced by the photon energy used in the ARPES measurements ($h\nu$ = 60 eV). In our experiments, we find that the spectral weight of the TSS decreases with increasing photon energy and becomes nearly undetectable above $\sim$10 eV, where bulk-band contributions dominate. Our results clearly demonstrate that the surface states are distinctly resolved only at low photon energies around 7 eV.

A similar photon-energy dependence has been reported in another antiferromagnetic topological insulator, MnBi$_2$Te$_4$~\cite{Hao2019-fo,Chen2019-qw,Baigeng_NC}, where higher photon energies predominantly probe bulk states with a large gap in the antiferromagnetic phase, while lower photon energies are required to clearly resolve the surface bands.\\

{\noindent\textbf {Supplementary Note 8: Evolution of the penetration depth of the topological surface state into the bulk across the magnetic transition}}
\vspace{0.3\baselineskip}

The Dirac gap is formed by the breaking of the symmetry composed of translational and time-reversal symmetries ($S$-symmetry), namely, by ferromagnetic order at the outermost surface. Therefore, the localization of the TSS wavefunction is crucial for the Dirac gap opening. To evaluate the surface localization, here, we estimate the penetration depth of the surface states into the bulk.

\begin{figure*}[h]
  \centering
  \includegraphics[width=0.9\textwidth]{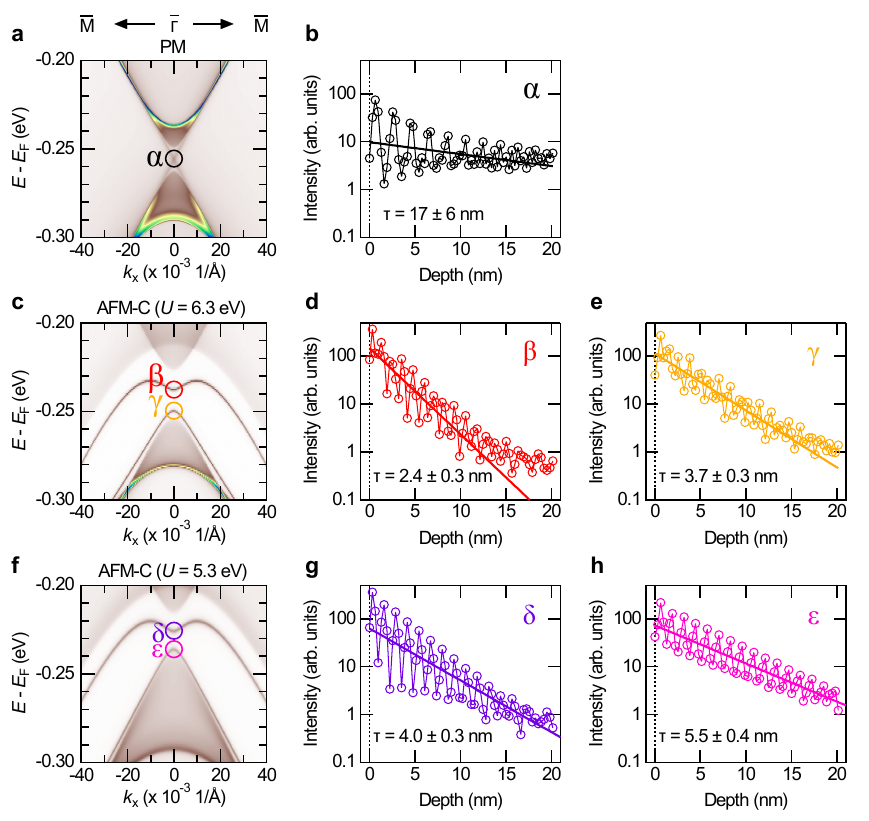}
    \label{sup_SPweight}
  \caption{
\textbf{Penetration depth analysis of the TSS across the magnetic transition in NdBi, based on DFT+$U$ calculations ($U=6.3$~eV).}
\textbf{a}, Surface band calculations of the paramagnetic (PM) phase showing the TSS at the marked energy-momentum point $\alpha$. 
\textbf{b}, Layer-resolved spectral weight of the TSS calculated from the surface (Depth=0~nm) to the bulk for $\alpha$. The decay length ($\tau$) was determined by fitting the data with an exponential decay function (solid line), yielding $\tau=17\pm 6$~nm.
\textbf{c}, Surface band calculations of the antiferromagnetic phase, Domain-C (AFM-C) using $U=6.3$~eV, with two representative points ($\beta$, $\gamma$) selected for comparison.
\textbf{d,e}, Layer-resolved spectral weight of the TSS calculated for $\beta$ and $\gamma$. The decay length was determined as $\tau=2.4\pm 0.3$~nm for $\beta$ and $\tau=3.7\pm 0.3$~nm for $\gamma$.
\textbf{f}, Surface band calculations of the antiferromagnetic phase, Domain-C (AFM-C) using $U=5.3$~eV, with two representative points ($\delta$, $\epsilon$) selected for comparison.  
\textbf{g,h},  Layer-resolved spectral weight of the TSS calculated for $\delta$ and $\epsilon$. The decay length was determined as $\tau=4.0\pm 0.3$~nm for $\delta$ and $\tau=5.5\pm 0.4$~nm for $\epsilon$.
    }
\end{figure*}

For this purpose, we calculate the layer-dependent spectral weight of TSS in the paramagnetic phase (PM) and the antiferromagnetic phase, Domain-C (AFM-C). We particularly focused on the energy-momentum states labeled as $\alpha$ (for PM) and $\beta$, $\gamma$ (for AFM-C) on the surface calculation maps in Fig.~\ref{sup_SPweight}a and c, respectively. 
The calculated results are shown in Fig.~\ref{sup_SPweight}b, d and e. The spectral weight significantly decreases inside the bulk in all states. To highlight the difference, we present the same data on a logarithmic scale.
It is now clear that the $\alpha$ state in the PM phase has a relatively long decay length, whereas the $\beta$ and $\gamma$ states in the AFM-C phase show more rapid decay inside the bulk. In Fig.~\ref{sup_SPweight}b, d and e, we perform an exponential fit to each dataset and estimate the penetration depth $\tau$. The estimated values of $\tau$ are explicitly indicated in each panel. Interestingly, the penetration depth in the PM phase is relatively long ($\sim$17 nm), whereas it gets much shorter (2 to 4 nm), reduced by a factor of $\sim$5-7, in the AFM-C phase. The significant reduction of $\tau$ across the magnetic transition originates from a qualitative change in the coupling between the surface and bulk states.
In the PM phase, the TSS overlaps with the bulk bands projected onto the surface, indicated as spectral continuum in Fig.~\ref{sup_SPweight}a. The consequent wavefunction mixing between TSS and bulk states allows the TSS to penetrate deep inside the bulk. In contrast, TSS in the AFM-C resides within the bulk band gap largely opened under antiferromagnetism, as demonstrated in Fig.~\ref{sup_SPweight}c. This strongly suppresses wavefunction mixing between the surface and bulk states. This reduced coupling results in the surface states being more localized at the surface without penetrating into the bulk in the AFM phase.

To test the dependence on the $U$ parameter, we performed another calculation for the antiferromagnetic phase, Domain-C using $U=5.3$~eV in Fig.~\ref{sup_SPweight}f (We used $U=6.3$~eV in the main manuscript). The calculated decay lengths of the upper and lower parts of TSS with this parameter are $4\sim6$~nm. This value confirms that the TSS in the AFM phase is more localized to the surface compared to the paramagnetic phase ($\sim17$ nm), regardless of the variation in $U$.
Regarding the surface termination, there is structurally only one surface termination along the (001) cleavage plane due to the NaCl structure. While the magnetic orientation of the topmost layer can be reversed, this simply inverts the sign of the $S_z$ spin texture and does not affect the decay length. Therefore, the surface termination effect is not present in this system.\\

{\noindent\textbf {Supplementary Note 9: Spin-resolved DFT calculations in the paramagnetic and antiferromagnetic phases}}
\vspace{0.3\baselineskip}

Here, we present spin-resolved DFT calculations, which are an effective means of distinguishing surface and bulk states, for both the paramagnetic and antiferromagnetic phases. 

Figure \ref{sup_spincalcPMAFM}a shows the surface spectral function (spin-integrated) for the PM phase. 
The TSS and bulk bands overlap in energy-momentum space, making them indistinguishable in spin-integrated spectra. This ambiguity, however, is resolved by considering spin polarization. Figure \ref{sup_spincalcPMAFM}b presents the corresponding spin-resolved spectral function. The TSS appears at the edge of the bulk continuum and exhibits a clear spin polarization. Although the bulk states exhibit weak spin polarization due to surface-induced inversion symmetry breaking, their spin polarization is significantly weaker and, importantly, of opposite sign to that of the TSS.

We further performed spin-resolved calculations for the antiferromagnetic (AFM) phase (Figs.~\ref{sup_spincalcPMAFM}c and \ref{sup_spincalcPMAFM}d). The results show that the sign of the spin polarization of the TSS remains unchanged across the AFM transition. In the AFM phase, the TSS is well separated from the bulk bands and can be clearly identified (see Figs. 3g and 3h in the main manuscript), providing a reliable benchmark for interpreting the experimental data. We find that the same spin polarization is observed in the PM phase, indicating that the feature assigned as the TSS in the PM phase originates from the same surface state. If the peaks were instead derived from bulk states, their spin polarization would exhibit the opposite sign according to our calculations; however, this is not observed. This directly rules out a bulk origin of the observed features.

The consistency of the spin polarization between the PM and AFM phases provides strong evidence that our 7 eV laser ARPES selectively probes the TSS even in the PM phase. In our measurements, the bulk contribution is strongly suppressed, likely reflecting a combination of intrinsically weaker bulk spectral weight and photoemission matrix element effects, which enhance the relative visibility of the surface state.
\\

{\noindent\textbf {Supplementary Note 10: Quantitative analysis of the spin polarization of the TSSs observed by spin-resolved laser ARPES}}
\vspace{0.3\baselineskip}

We present a quantitative analysis of the spin polarization as a function of energy at $\pm k$ in Fig.~\ref{sup_spinpol}. The error bars reflect the statistical uncertainty derived from counting statistics. The overall energy and momentum resolutions are described in the Methods section.\\

{\noindent\textbf {Supplementary Note 11: Determination of the topological Dirac gap size from laser-ARPES}}
\vspace{0.3\baselineskip}

In the main text, we report that the topological Dirac gap is $17\pm2$~meV. Here, we present a fitting analysis of the relevant spectrum to determine the gap size and its associated uncertainty.

\begin{figure*}[h]
  \centering
  \includegraphics[width=0.5\textwidth]{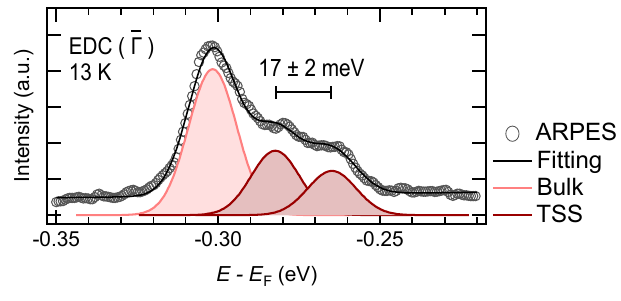}
    \label{sup_EDCgamma}
  \caption{
\textbf{Quantitative determination of the topological Dirac gap size at the $\bar{\Gamma}$ point.}
Energy distribution curve (EDC) at the $\bar{\Gamma}$ point measured at 13 K. The EDC was fitted using a combination of three Gaussian functions with a linear background. To stabilize the fitting, the widths of the three peaks were constrained to be equal.
 }
\end{figure*}

\begin{figure*}[h]
  \centering
  \includegraphics[width=0.9\textwidth]{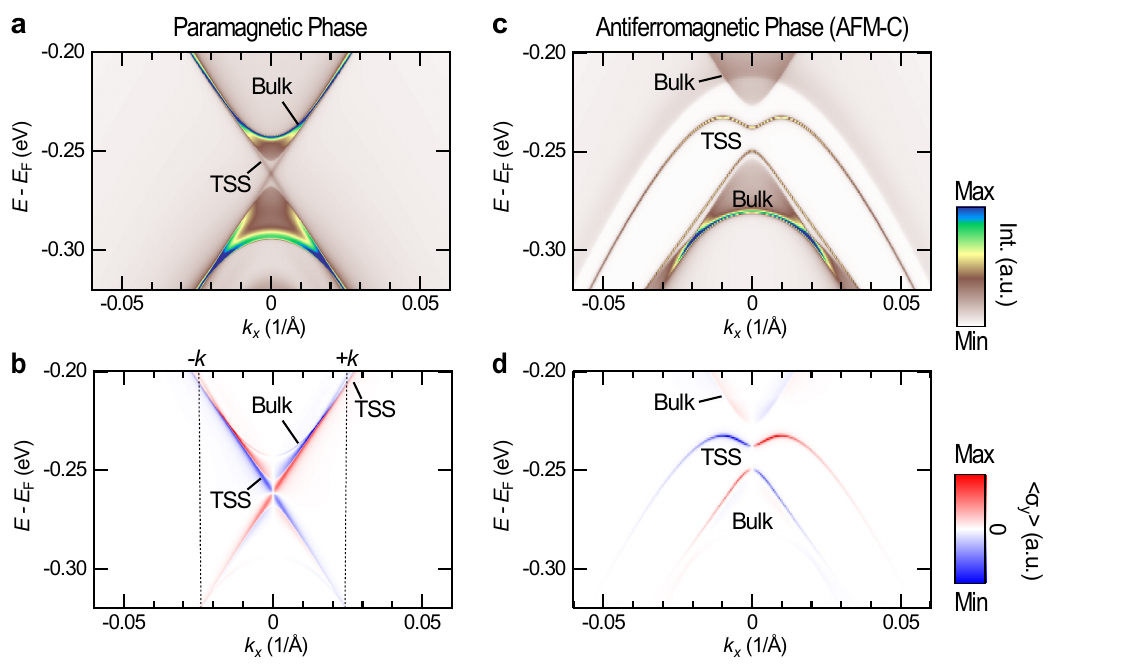}
    \label{sup_spincalcPMAFM}
  \caption{
\textbf{Spin-resolved DFT calculations of the electronic structure in the paramagnetic and antiferromagnetic phases.}
\textbf{a}, Spin-integrated spectral weight map for the paramagnetic (PM) phase. The topological surface states (TSS) and bulk bands overlap in energy-momentum space.
\textbf{b}, Spin-resolved spectral weight map for the paramagnetic phase. The color scale represents the expectation value for the $y$ component of spin. The TSS is clearly distinguishable by its spin polarization, which has the opposite sign compared to that of the bulk bands. The dashed lines indicate the momentum cuts corresponding to the spin-resolved ARPES measurements presented in Fig.~3g of the main text. 
\textbf{c,d}, Spin-integrated and spin-resolved spectral weight maps for the antiferromagnetic phase (Domain-C), respectively. The calculations in c and d are the same as those used in Fig.~3i of the main text.
    }
\end{figure*}

\begin{figure*}[h]
  \centering
  \includegraphics[width=0.9\textwidth]{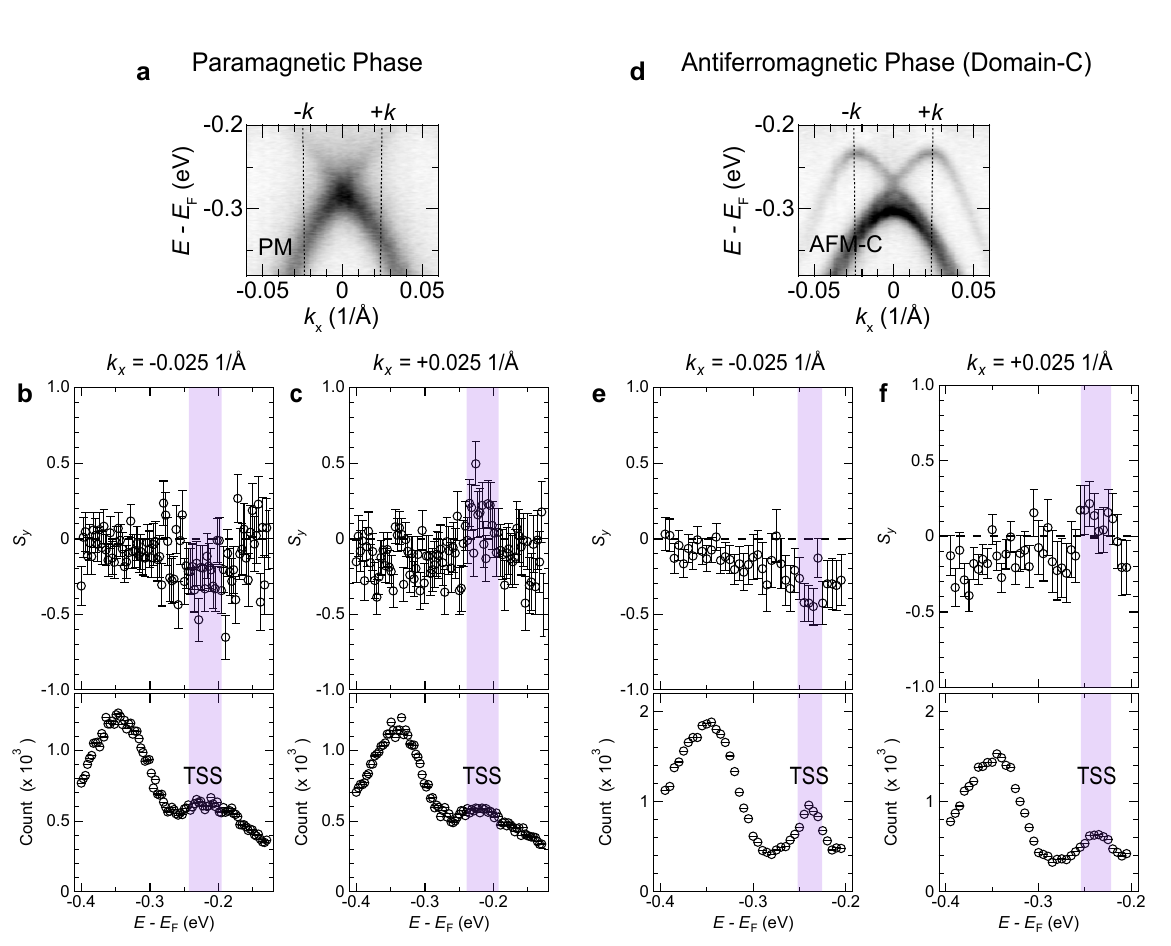}
    \label{sup_spinpol}
  \caption{
\textbf{Spin polarization analysis of the TSS in the paramagnetic and antiferromagnetic phases.}
\textbf{a}, ARPES intensity map along the $\bar{\Gamma}$--$\bar{\rm M}$ direction in the paramagnetic (PM) phase. The dashed vertical lines indicate the momentum positions ($\pm k$) where spin-resolved ARPES measurements were performed. 
\textbf{b,c}, Spin polarization (upper panel) and spin-integrated spectra (bottom panel) at $-k$ and $+k$ for the PM phase, respectively.
\textbf{d}, ARPES intensity map in the antiferromagnetic phase, Domain-C (AFM-C).
\textbf{e,f}, Spin polarization (upper panel) and spin-integrated spectra (bottom panel) at $-k$ and $+k$ for the AFM-C phase, respectively. The purple shaded regions highlight the spectral peaks originating from the topological surface states (TSS).
    }
\end{figure*}

Figure~\ref{sup_EDCgamma} shows the EDC at the $\bar{\mathrm{\Gamma}}$ point of the AFM-C domain (Fig.~3d in the main manuscript), together with the fitting curve.
We fit the EDC using a combination of three Gaussian functions with a linear background. This estimates the gap size to be $17\pm2$~meV.
To assess reproducibility, we repeated the measurements on three independent sample cleaves and consistently obtained similar spectral features and gap magnitudes, which ensures the reliability of our results. The uncertainty was determined from the fitting procedure.
\\

{\noindent\textbf {Supplementary Note 12: Critical scaling of the hybridization gap with temperature}}
\vspace{0.3\baselineskip}

\begin{figure*}[h]
  \centering
  \includegraphics[width=0.7\textwidth]{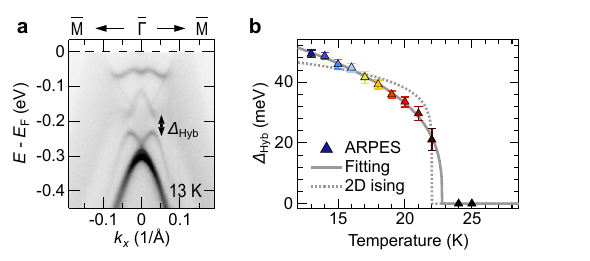}
    \label{sup_beta}
  \caption{
\textbf{Critical scaling of the hybridization gap $\Delta_{\mathrm{Hyb}}(T)$.}
\textbf{a}, ARPES intensity map of AFM-C domain at 13 K.
\textbf{b}, Temperature dependence of $\Delta_{\mathrm{Hyb}}$. The solid gray curve is a power-law fit: $\Delta_{\mathrm{Hyb}} \propto (1 - T/T_{\mathrm{N}})^\beta$ with $\beta$ = 0.32(6) and $T_{\mathrm{N}}$ = 22.8(7) K. For comparison, a fit using the 2D Ising exponent ($\beta$ = 1/8) is shown as a gray dotted line. The extracted exponent is consistent with the 3D Ising universality class.
    }
\end{figure*}

Below the $T_{\mathrm{N}}$, the TSS and the SS1 hybridize to form a gap, denoted as $\Delta_{\mathrm{Hyb}}$ (Fig.~\ref{sup_beta}a). The evolution of the hybridization gap is shown in Fig.~\ref{sup_beta}b. We found that the extracted temperature dependence of $\Delta_{\mathrm{Hyb}}$ is well described by a power-law fit $\Delta_{\mathrm{Hyb}} \propto (1 - T/T_{\rm N})^\beta$, with $\beta = 0.32(6)$ and $T_{\rm N}=22.8(7)$~K. The obtained critical exponent $\beta$ closely matches the theoretical magnetic order parameter for a 3D Ising model~\cite{Guida1998-ci}, whereas it deviates significantly from that of a 2D Ising model (indicated by the dotted line in Fig.~\ref{sup_beta}b). 

These results demonstrate that the surface magnetism evolves in tandem with the bulk magnetic order, at least on a freshly cleaved surface. For the fitting procedure, the surface $T_{\mathrm{N}}$ was treated as a free parameter. The slight deviation ($\sim 1.2$~K) from the bulk $T_{\mathrm{N}}$ (24 K) is likely attributable to minor experimental factors such as a thermal gradient between the sample holder and the sample surface, calibration offset of the thermometer, or potentially a small intrinsic reduction of $T_{\mathrm{N}}$ at the surface. In any case, this small difference does not affect our main conclusions.\\

{\noindent\textbf {Supplementary Note 13: Quantitative analysis of the time evolution of the hybridization gap}}
\vspace{0.3\baselineskip}

\begin{figure*}[h]
  \centering
  \includegraphics[width=0.9\textwidth]{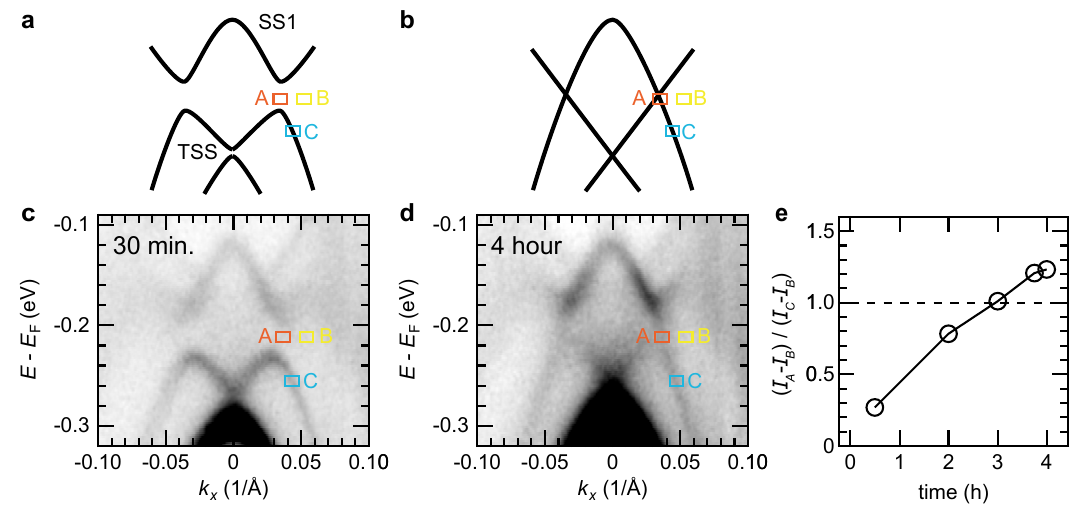}
    \label{sup_timedep}
  \caption{
\textbf{Effect of surface contamination on the electronic structure.}
\textbf{a,b}, Schematics of the surface states for a clean surface (\textbf{a}) and a contaminated surface (\textbf{b}). On the clean surface, a gap $\Delta_{\mathrm{Hyb}}$ is formed between the topological surface state (TSS) and the surface state (SS1), whereas the gap collapses upon surface contamination. 
\textbf{c,d}, ARPES intensity maps measured at 30 minutes (\textbf{c}) and 4 hours (\textbf{d}) after cleavage.
\textbf{e}, Time evolution of the photoelectron intensity within the gap. The intensity was calculated from the integrated spectral intensity of the A-C rectangular regions marked in \textbf{c} and \textbf{d}.
    }
\end{figure*}

To quantify the time evolution of the hybridization gap, we analyzed the ratio of the spectral intensity within the gap to that of the surface state, as shown in Fig.~\ref{sup_timedep}e, where the horizontal axis represents the elapsed time after cleavage. The data show a monotonic increase in the in-gap intensity over time, quantitatively confirming progressive spectral filling that leads to gap closure after $\sim$4 hours. The uncertainties in the intensity ratio were estimated from the statistical variation of the integrated spectral weight and are included in Fig.~\ref{sup_timedep}e.\\

\noindent\textbf{Supplementary Note 14: Temperature and pressure stability of the experimental setup}
\vspace{0.3\baselineskip}

\begin{figure*}[h]
  \centering
  \includegraphics[width=0.9\textwidth]{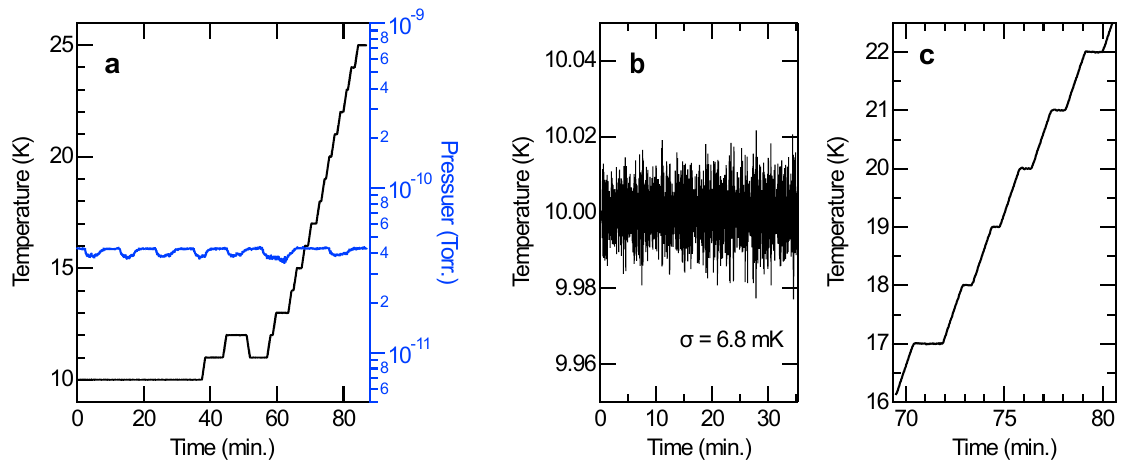}
    \label{sup_temphist}
  \caption{
\textbf{Stability of sample temperature and vacuum conditions.}
\textbf{a}, Time evolution of the sample temperature (left) and vacuum pressure (right) under PID control.
\textbf{b}, Temperature stability maintained at 10 K. The standard deviation over a 30-minute period is 6.8 mK.
\textbf{c}, Temperature profile during a controlled ramp at a rate of 1 K/min.
    }
\end{figure*}

To confirm the temperature and pressure stability of our experiments, we conducted independent tests over a period of one hour. Fig.~\ref{sup_temphist}a demonstrates that the base pressure remains highly stable at $<5\times 10^{-11}$~Torr over more than one hour at low temperatures, representing the typical pressure history during our measurements. Regarding thermal stability, the temperature was controlled using a PID controller. As shown in Fig.~\ref{sup_temphist}b, the standard deviation over a 30-minute period was 6.8 mK, confirming negligible thermal drift within the experimental time window. Furthermore, Fig.~\ref{sup_temphist}c confirms that the temperature sweep rate of 1 K/min used in our experiments does not induce any pressure fluctuations.\\

\renewcommand{\baselinestretch}{1.15}

\clearpage
%

